\documentclass[%
 reprint,
 amsmath,amssymb,
 aps,
]{revtex4-1}

\usepackage{graphicx}
\usepackage{dcolumn}
\usepackage{bm}

\usepackage{hyperref}
\usepackage{fancyhdr}
\usepackage{latexsym}
\usepackage{graphicx,color}

\usepackage{shadow,epsf,amsthm,amssymb,amsmath}

\usepackage{enumerate}

\usepackage{hvdashln}                           
\usepackage{setspace}                           

\usepackage{lipsum}

\newtheorem{definitionenv}{Definition}
\newtheorem{lemmaenv}[definitionenv]{Lemma}
\newtheorem{theoremenv}[definitionenv]{Theorem}
\newtheorem{corollaryenv}[definitionenv]{Corollary}
\newtheorem{propositionenv}[definitionenv]{Proposition}
\newtheorem{conjectureenv}[definitionenv]{Conjecture}
\newtheorem{exampleenv}{Example}
\newtheorem{app-lemmaenv}[section]{Lemma}

\newenvironment{definition}{\begin{definitionenv}\rm}{\end{definitionenv}}
\newenvironment{lemma}{\begin{lemmaenv}\rm}{\end{lemmaenv}}
\newenvironment{theorem}{\begin{theoremenv}\rm}{\end{theoremenv}}
\newenvironment{corollary}{\begin{corollaryenv}\rm}{\end{corollaryenv}}
\newenvironment{example}{\begin{exampleenv}\rm}{\end{exampleenv}}
\newenvironment{proposition}{\begin{propositionenv}\rm}{\end{propositionenv}}
\newenvironment{conjecture}{\begin{conjectureenv}\rm}{\end{conjectureenv}}
\newenvironment{app-lemma}{\begin{app-lemmaenv}\rm}{\end{app-lemmaenv}}

\newcommand{\bd}{\begin{definition}}
\newcommand{\ed}{\end{definition}}
\newcommand{\bl}{\begin{lemma}}
\newcommand{\el}{\end{lemma}}
\newcommand{\elp}{\hspace*{\fill} $\Box$
                 \end{lemma}}
\newcommand{\bt}{\begin{theorem}}
\newcommand{\et}{\end{theorem}}
\newcommand{\etp}{\hspace*{\fill} $\Box$
                 \end{theorem}}
\newcommand{\bc}{\begin{corollary}}
\newcommand{\ec}{\end{corollary}}
\newcommand{\ecp}{\hspace*{\fill} $\Box$
                 \end{corollary}}
\newcommand{\bcj}{\begin{conjecture}}
\newcommand{\ecj}{\end{conjecture}}

\newcommand{\be}{\begin{example}}
\newcommand{\ee}{\end{example}}
\newcommand{\eep}{\hspace*{\fill} $\Box$
                 \end{example}}
\newcommand{\bp}{\begin{proposition}}
\newcommand{\ep}{\end{proposition}}
\newcommand{\epp}{
                 \end{proposition}}

\newcommand{\bra}[1]{\langle#1|}
\newcommand{\ket}[1]{|#1\rangle}

\newcommand{\tr}{\mbox{tr}}
\newcommand{\wt}[1]{\mbox{wt}(#1)}

\begin{document}

\preprint{APS/123-QED}

\title{Entanglement-Assisted Quantum Error-Correcting Codes with Imperfect Ebits}

\author{Ching-Yi Lai}
\email{laiching@usc.edu}
\author{Todd A. Brun}%
\email{tbrun@usc.edu}
 \affiliation{%
 Electrical Engineering Department, University of Southern California, Los Angeles, California, USA  90089.\\
}%

%
%

\date{\today}

\begin{abstract}
The scheme of entanglement-assisted quantum error-correcting (EAQEC) codes assumes that
the ebits of the receiver are error-free.
In practical situations, errors on these ebits are unavoidable, which diminishes the error-correcting ability of these codes.
We consider  two different versions of this problem.
We first show that any (nondegenerate) standard stabilizer code can be transformed into an EAQEC code that can correct errors on the qubits of both sender and receiver.
These EAQEC codes are equivalent to standard stabilizer codes, and hence the decoding techniques of standard stabilizer codes can be applied.
Several EAQEC codes of this type are found to be optimal.
In a second scheme, the receiver uses a standard stabilizer code to protect the ebits, which we call a ``combination code."
The performances of different quantum codes are compared in terms of the channel fidelity over the depolarizing channel.
We give a formula for the channel fidelity over the depolarizing channel (or any Pauli error channel), and show that it can be efficiently approximated by a Monte Carlo calculation.
Finally, we discuss the tradeoff between performing extra entanglement distillation and applying an EAQEC code with imperfect ebits.
\end{abstract}

\maketitle

\section{Introduction}

The theory of quantum error correction is important for both quantum computation and quantum communication \cite{Shor95,Ste96a,Ste96b,EM96,BDSW96,KL97}.
Quantum stabilizer codes are the most extensively studied quantum codes \cite{CRSS97,Got97}, and  have the advantage that
their properties can be analyzed using group algebra.
Quantum stabilizer codes are closely related to classical linear codes, and can be obtained by the CRSS and CSS code constructions from  weakly self-dual classicalcodes \cite{CS96,Ste96,CRSS97,CRSS98,Got97,NC00}.
 When entanglement between sender and receiver is available, a new error correction scheme becomes possible: entanglement-assisted quantum error-correction.
 This coding scheme (EAQEC codes) has the advantage that it allows any classical linear code, not necessarily weakly self-dual, to be transformed into a quantum code \cite{BDM06}.
In addition, EAQEC codes can increase  both the transmission rate and error-correcting ability \cite{LB10,WH10}.
Also, some problems or limitations in quantum LDPC codes and turbo codes can be solved using EAQEC codes \cite{WH10,HBD08}.

In EAQEC codes \cite{BDM06}, it is assumed that the sender (Alice) and the receiver (Bob) share some pairs of qubits in maximally-entangled states (ebits) before communication, and the quits on Bob's side are subject to no error.
The quantum codes are designed to cope with the noisy channel $\mathcal{N}_A$ that Alice uses to communicate with Bob.
The properties of EAQEC codes in this case are studied in \cite{BDM06,LB10,LBW10}.
However, noise  (such as storage errors) can occur on Bob's ebits in practical situations, which is believed to degrade the performance of the quantum codes.

Assume the errors occurring on Bob's qubits are described by a noise process $\mathcal{N}_B$.
Wilde and Hsieh addressed this question with a channel-state coding protocol in quantum Shannon theory and determined the channel capacity
when entanglement is not perfect \cite{WH10a}.
They also performed simulations of entanglement-assisted quantum turbo codes with the depolarizing channel when Bob's ebits also suffer errors \cite{WH10}.
Wilde and Fattal simulated the performance of an entanglement-assisted Steane code for fault tolerance \cite{WF10}.


In this article, we discuss two coding schemes to handle the problem when the ebits of Bob are not perfect.
Shaw et al. described  a six-qubit EAQEC code with one ebit that is equivalent to  Steane's seven-qubit code, and can correct a single error on either Alice's or Bob's qubits \cite{SWOKL08}.
The entanglement-assisted Steane code, constructed by Wilde and Fattal, is also equivalent to Steane's seven-qubit code \cite{WF10}.
Similarly, Bowen's entanglement-assisted code \cite{Bowen02} is equivalent to the five-qubit code \cite{BDSW96,LMPZ96} and can correct an error on one of Bob's qubits.
These three examples motivate the following idea: there are EAQEC codes that are equivalent to standard stabilizer codes, and hence can correct errors on both Alice's and Bob's sides.
We show how to obtain an EAQEC code from a (nondegenerate) stabilizer code.
Several EAQEC codes from this scheme are found to be \emph{optimal}.
We say a quantum code is optimal if the minimum distance of this code achieves an upper bound for fixed numbers of information qubits and physical qubits.
These EAQEC codes will have better performance than their equivalent stabilizer codes when
the storage error rate is less than the channel error rate.

In the second scheme, Alice uses an EAQEC code  to encode her information qubits and  Bob uses a standard stabilizer code to protect his halves of the ebits.
The combination of an EAQEC code and a stabilizer code is called a combination code, and it can be treated either as a single stabilizer code, or by using two sequential decoders.
EAQEC codes that are not equivalent to standard stabilizer codes generally have higher error-correcting ability on Alice's qubits and are suitable for this scheme.

Minimum distance of a stabilizer code is used as a measure of how good a code is without considering the details of the noisy channel model.
However, minimum distance might not always be the best measure, for a quantum code may be able to correct many error operators of weight higher than that indicated by the minimum distance.
In particular, there is no general definition of minimum distance for the variant coding schemes in this article.
A perhaps more suitable merit function is the \emph{channel fidelity} \cite{Schumacher96,RW06}, which compares the similarity of the modified quantum state with the original quantum state.
However, the calculation of the channel fidelity depends on the channel and has an exponentially increasing complexity.
We derive a formula for the channel fidelity of a quantum stabilizer code over the depolarizing channel, which facilitates its computation.
The channel fidelity also can be well approximated by a lower bound when the depolarizing rate is small.
Furthermore, Monte Carlo methods can often efficiently approximate  the channel fidelity \cite{MU49}.

Another natural question arises in EAQEC codes.
The perfect entanglement shared between sender and receiver will in practice be generated
from a process of entanglement distillation \cite{BBPSSW96PhysRevLett.76.722,BDSW96} or a breeding protocol \cite{LD07PhysRevA.75.010303}.
It is known that entanglement distillation  with one-way classical communication is equivalent to
a quantum error-correcting code \cite{BDSW96}.
Since we can also communicate using an EAQEC code that is robust to  imperfect ebits, we discuss whether it is always necessary to do
entanglement distillation before communication, and how much.

This paper is structured as follows.
Basics of stabilizer codes and EAQEC codes are reviewed in the next section.
Basic criteria for an EAQEC code to be capable of correcting errors on Bob's qubits are analyzed in Section \ref{sec:syndrome_representatives}.
We discuss the first coding scheme with imperfect ebits in Section \ref{sec:standard_equivalent}, and the second scheme in Section \ref{sec:two_encoder}.
The formula for the channel fidelity over the depolarizing channel is derived in Section \ref{sec:channel_fidelity},
and the channel fidelities for quantum stabilizer codes and EAQEC codes are  given in Subsection \ref{subsec:fc_formula} and \ref{subsec:fc_formula_EAQEC}, respectively.
We discuss Monte Carlo simulations for the channel fidelity in Subsection \ref{subsec:monte_carlo}.
In Section  \ref{sec:comparison}, we compare the performances of different coding schemes in terms of the channel fidelity.
Entanglement distillation is discussed in Section \ref{sec:Entanglement Distillation}.
The conclusions follow in Section \ref{sec:discussion}.

\section{Basics}
\label{sec:basics}

Let $\mathcal{H}$ be the state space of a single qubit.
Suppose Alice sends a $k$-qubit state $|\psi\rangle$ to Bob
by using an $[[n,k,d]]$ quantum stabilizer code that encodes the $k-$qubit information state $\ket{\psi}$ in  a
$2^k$-dimensional subspace of the $n$-qubit state space $\mathcal{H}^{\otimes n}$, fixed by a stabilizer group $\mathcal{S}$  with some minimum distance $d$.
The stabilizer group $\mathcal{S}$ is an Abelian subgroup of the $n$-fold Pauli group $\mathcal{G}_n$, with $n-k$ generators  $g_1,g_2,\cdots, g_{n-k}$, and does not contain the negative identity operator $-I$.
Suppose $U_E$ is an $n$-fold unitary \emph{Clifford encoder}, which leaves the $n$-fold Pauli group $\mathcal{G}_n$ invariant under conjugation.
The encoded state is $ U_E\left(\ket{\psi}\ket{0}^{\otimes n-k}\right)$.
For convenience, let the stabilizer generators be $g_i= U_E Z_{i+k} U_E^{\dag}$ for $i=1, \cdots, n-k$, where the subscript $i+k$ of the Pauli operator means that the operator is on the $(i+k)$-th qubit.
The error-correction condition for stabilizer codes says that $\{E_i\}$ is a set of correctable error operators in $\mathcal{G}_n$ if
$E_i^{\dag}E_j\notin \mathcal{N}(\mathcal{S})\backslash \mathcal{S}$ \cite{Got97,NC00},
where $\mathcal{N}(\mathcal{S})$ is the normalizer group of $\mathcal{S}$ in $\mathcal{G}_n$.
(Since the the overall phase of a quantum state is not important, we consider errors of the form $M_1\otimes \cdots \otimes M_n$, where $M_j\in \{I,X,Y,Z\}$ for $j=1,\cdots,n$.)
This implies the definition of the minimum distance $d$ of a stabilizer code to be the minimum weight of any element in $\mathcal{N}(\mathcal{S})\backslash \mathcal{S}$,
where the weight $\mbox{wt}(g)$ of $g\in \mathcal{G}_n$
 is  the number of components of $g$ that are not equal to the identity.

After Bob receives the noisy quantum state, he does the following three steps to recover the information state $\ket{\phi}$: syndrome measurement, correction, and decoding.
He first applies a series of projective measurements with projectors \[
    P_s=\prod_{j=1}^{n-k}\frac{I+(-1)^{s_j} g_j}{2}
\] on the  output state of the noisy channel.
Here, $s=s_1 s_2 \cdots s_{n-k}$ is a binary $(n-k)$-tuple that represents the \emph{error syndrome}.
The error syndrome $s$ of an error operator $E$ has $s_j=0$ if $E$ commutes with $g_j$ and $s_j=1$ otherwise.
We sometimes represent $s$ as a number with binary form $s_1 \cdots s_{n-k}$.
Note that the $P_s$'s  are orthogonal to each other, and $P_0$ is the projector on the code space, that is,
\[
P_0 U_E\left(\ket{\psi}\otimes \ket{0}^{\otimes n-k}\right) = U_E\left(\ket{\psi}\otimes \ket{0}^{\otimes n-k}\right)
\]
for any $k$-qubit state $\ket{\psi}$.
Given a stabilizer group  $\mathcal{S}$, there are $2^{n-k}$ distinct error syndromes. 
For each nonzero error syndrome $s$, we choose a Pauli operators $E_s$ (not in  $\mathcal{S}$), whose error syndrome is $s$;
for the error syndrome $s=0$, we choose $E_0=I$. The error operators $E_s$ are called \emph{syndrome representatives}.
If the measurement result is $s$,  the correction operator $C_s = E_s$  is applied, followed by the decoding unitary operator $U_E^{\dag}$.
Finally, Bob throws away the ancilla qubits, which is the same as applying a partial trace over the ancilla qubits.
We define a set $T$ containing the syndrome representatives $\{E_s\}$. 
Then $|T|=2^{n-k}$ and $T$ is a set of correctable error operators.
Note that the choice of $T$ determines the decoding process.
($\mathcal{S}$ determines the encoding process.)
In fact, we have many more correctable error operators than $T$.
For $g\in \mathcal{S}$, the operation of $E_s g$ and $E_s$ on the encoded state are the same
($
E_s g U_E\ket{\psi}\ket{0}^{\otimes n-k}=  E_s U_E\ket{\psi}\ket{0}^{\otimes n-k}
$)
 and can be corrected by the the same correction operator $C_s$.
 The error operator $E_sg$ is called a \emph{degenerate error} of $E_s$.
The set $T\times \mathcal{S}= \{ hg: h\in T, g\in \mathcal{S} \}$ is a correctable set of error operators that satisfies the error correction condition.
\bl
For a given $T$, the error operators in $T\times \mathcal{S}$ are correctable and are the only correctable error operators.
\el
\noindent Therefore, we have a total of $4^{n-k}$ correctable $n$-fold Pauli operators.
Note that a code can have different and inequivalent sets $T$.

In this article, error processes are modeled by the depolarizing channel $T_p^{\otimes n}$  independently  operating on $n$ qubits, where
\[
T_p(\rho)= (1-\frac{3}{4}p)\rho + \frac{p}{4}(X\rho X+ Y\rho Y+ Z\rho Z),
\]
with depolarizing rate $p$ for  $0\leq p\leq 1$ and  $\rho$ the density operator of a single qubit.
Since the operation elements of the depolarizing channel $\mathcal{T}_p$ are $\sqrt{1-\frac{3}{4}p}I,$ $\sqrt{\frac{1}{4}p}X,$ $\sqrt{\frac{1}{4}p}Y,$ and $\sqrt{\frac{1}{4}p}Z$,
the operation elements of  $\mathcal{T}_p^{\otimes n}$  are $\{\sqrt{p_i} E_i\}$,
where $E_i$ is a Pauli operator in the $n$-fold Pauli group $\mathcal{G}_n$ 
and $p_i$ is the probability that error $E_i$ happens.
If $E_i$ is of weight $w$, then $p_i$ is
\begin{align}
    q_w\triangleq (1-\frac{3}{4}p)^{n-w}(\frac{1}{4}p)^w. \label{eq:probability_def}
\end{align}

Now assume  $c$ maximally-entangled states $|\Phi_{+}\rangle
^{AB}=\frac{1}{\sqrt{2}}\left(  |00\rangle+|11\rangle\right)$ are shared between Alice and Bob.
Suppose Alice uses an $n$-fold Clifford encoder $U$ to encode a $k$-qubit state $|\psi\rangle$ in $n$ physical qubits (including the $c$ halves of the ebits on Alice's side) and then sends it to Bob.
This is called an  $[[n,k,d;c]]$ EAQEC code for some minimum distance $d$.
 The encoded state  is
\[
(U^{A}\otimes I^{B})\left(  |\psi\rangle \otimes(|\Phi
_{+}\rangle^{AB})^{\otimes c} \otimes \ket{0}^{\otimes n-k-c} \right)  ,
\]
where the superscript $A$ or $B$ indicates that the operator acts on the
qubits of Alice or Bob, respectively.
Let ${g'}_j= U Z_j U^{\dag}$ and ${h'}_j=UX_jU^{\dag}$ for $j=1, \cdots, n$.
The encoded state is stabilized by ${g'}_i^A\otimes Z_i^B$ for $i=k+1,\cdots, k+c$,
${h'}_i^A\otimes X_i^B$ for $i=k+1,\cdots, k+c$, and ${g'}_i^A\otimes I^B$ for $i=k+c+1,\cdots, n$.
These ${g'}_i$'s and ${h'}_j$'s are called \emph{simplified} stabilizer generators.
(We will omit the superscripts $A, B$ as there is no ambiguity.)
Note that ${g'}_i$ and ${h'}_i$ anti-commute with each other and they commute with
all other generators ${g'}_j$ and ${h'}_j$ for $j \neq i$.
Thus  ${g'}_i$ and ${h'}_i$ are symplectic partners for $i=1, \cdots, k+c$.
An EAQEC code is defined by the simplified stabilizer group $\mathcal{S}'$ of the encoded state:
\[\mathcal{S}'= \langle {{g'}}_{k+1}, \cdots, {g'}_{k+c},
{h'}_{k+1}, \cdots, {h'}_{k+c}, {g'}_{k+c+1}, \cdots, {g'}_{n} \rangle.\]
The symplectic subgroup of $\mathcal{S}'$  is
$\mathcal{S}_S=$ $\langle {g'}_{k+1}$, $\cdots,$ ${g'}_{k+c}$, ${h'}_{k+1}$, $\cdots,$ ${h'}_{k+c}\rangle$,
and  the isotropic subgroup of $\mathcal{S}'$ is
$\mathcal{S}_I =$ $\langle g_{k+c+1,}$, $\cdots,$ ${g'}_{n}\rangle$.
$\mathcal{S}'= \mathcal{S}_S\times \mathcal{S}_I$,
so that $\mathcal{S}'$ is generated by the generators of $\mathcal{S}_S$ and $\mathcal{S}_I$.
The minimum distance of the EAQEC code is the minimum weight of any element in $\mathcal{N}(\mathcal{S}')\backslash \mathcal{S}_I$ \cite{BDM06}.
The decoding process of an EAQEC code is similar to that of a standard stabilizer code.

\section{Determining the Syndrome Representatives for EAQEC Codes} \label{sec:syndrome_representatives}

For a channel with low error rate, we would like to choose the set of syndrome representatives $T$ to contain lower-weight error operators, since these errors
are more likely to occur.
A simple algorithm to define a set $T$ is as follows:
\begin{enumerate}[(a)]
    \item Let $T=\emptyset$ and  $T'=\emptyset$
    \item Find an error operator $E\in \mathcal{G}_n\setminus (T\cup T')$ with the lowest weight, and compute its syndrome $s$.
    \item If there is no $E_s\in T$, set $E_s\leftarrow E$ and $T\leftarrow T\cup \{E_s\}$.
    \item If there is an $E_s\in T$, set  $T'\leftarrow T'\cup \{E\}$.
    \item If $|T|<2^{n-k}$, go to (b). Else, output $T$.
\end{enumerate}
This algorithm finds a set $T$ of minimum weight, that is, it minimizes the quantity
\[
\sum_{E\in T} \mbox{wt}(E).
\]

In the usual paradigm of EAQEC codes, it is assumed that Bob's qubits suffer no error.
However, this assumption might not be true in practice.
Suppose that Alice uses a noisy channel $\mathcal{N}_A$ to communicate with Bob and Bob's ebits suffer from a storage error channel $\mathcal{N}_B$.
Assume both $\mathcal{N}_A$ and $\mathcal{N}_B$ are  depolarizing channels.
Let $p_a$ and $p_b$ be the depolarizing rate of  $\mathcal{N}_A$ and $\mathcal{N}_B$, respectively.
Define
\begin{align}\label{eq:probability_def_Na}
q_w\triangleq \left(1-\frac{3}{4}p_a\right)^{n-w}\left(\frac{1}{4}p_a\right)^w,
\end{align}
for $w=0, \cdots, n$, and
\begin{align} \label{eq:probability_def_Nb}
r_{w'}\triangleq \left(1-\frac{3}{4}p_b\right)^{c-w'}\left(\frac{1}{4}p_b\right)^{w'}.
\end{align}
for $w'=0, \cdots, c$.
An error operator $E_A \otimes E_B$ of $\mathcal{N}_A \otimes \mathcal{N}_B$ occurs with probability $q_{w_a}r_{w_b}$, where  $w_a=\wt{{E_A}}$ and $w_b=\wt{{E_B}}$.

\begin{figure}
\begin{center}
\includegraphics[height=6.5cm]{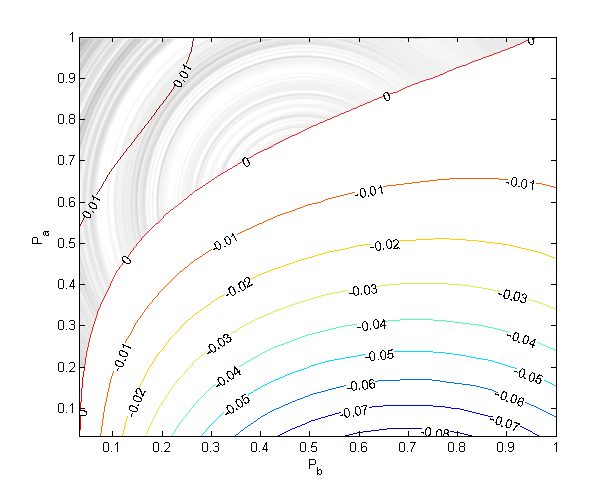} 
{\caption{(Color online) Contour plot of $q_2r_0-q_0r_1$.
The shaded region is where $q_2r_0-q_0r_1>0$.
}\label{fig:q0r1_q2r0}}
\end{center}
\end{figure}

To correct some errors on Bob's qubits, we have to design a quantum code
such that these errors are either syndrome representatives, or degenerate errors of other correctable errors.
It is more complicated to determine the syndrome representatives in the case of EAQEC codes, since the error probabilities are different on Alice's and Bob's qubits.
For an EAQEC code to correct some errors on Bob's qubits,
we must sacrifice some of its ability to correct channel errors.
For example, consider Bowen's $[[3,1,3;2]]$ code with the following stabilizer generators:
\begin{align} \label{eq:Bowen's_code}
\begin{array}{ccc|cc}
X&Z&Z&X&I\\
Z&Z&X&I&X\\
Z&Y&Y&Z&I\\
Y&Y&Z&I&Z.
\end{array}
\end{align}
Alice's qubits and Bob's qubits are on the left and the right of the vertical line, respectively.
The error operators $X_4$ and $Y_1Y_2$ have the same error syndrome.
$X_4$ is an error operator on Bob's side with probability
\[
q_0r_1=\left(1-\frac{3}{4}p_a\right)^{3}\left(1-\frac{3}{4}p_b\right)\left(\frac{1}{4}p_b\right).
\]
If Bob's qubits are error free, the code can correct the weight-2 error $Y_1Y_2$ on Alice's side, which occurs with probability
\[
q_2r_0=\left(1-\frac{3}{4}p_a\right)\left(\frac{1}{4}p_a\right)^{2}\left(1-\frac{3}{4}p_b\right)^2.
\]
We can instead choose to correct $X_4$ if $q_0r_1>q_2r_0$. 
This is a tradeoff between correcting channel errors or storage errors.
We plot $q_2r_0-q_0r_1$ as a function of $p_a$ and $p_b$ in Fig. \ref{fig:q0r1_q2r0}.
It can be seen that  $Y_1Y_2$ is a more likely error than $X_4$ when $p_b$ is small or $p_a$ is large.

To sum up, the set of syndrome representatives $T$ should be chosen to minimize
\begin{align*}
\sum_{E\in T} \mbox{Pr}(E).
\end{align*}
We will show that this criterion leads to high channel fidelity in Section \ref{sec:channel_fidelity}.
We propose two coding schemes in the next two sections.


\section{EAQEC Codes that are Equivalent to Standard Stabilizer Codes}
\label{sec:standard_equivalent}

The stabilizer generators of the five qubit code \cite{NC00} are:
\begin{align*}
\begin{array}{ccccc}
X&Z&Z&X&I\\
I&X&Z&Z&X\\
X&I&X&Z&Z\\
Z&X&I&X&Z
\end{array}
\end{align*}
After some row operations, these are equivalent to  the stabilizer generators of  Bowen's $[[3,1,3;2]]_{AB}$ EAQEC code in (\ref{eq:Bowen's_code}).
Thus Bowen's quantum code is equivalent to the $[[5,1,3]]$ standard stabilizer code, and
has the same error-correcting ability on Bob's halves of the ebits as on Alice's qubits.
We use the subscript $AB$ to indicate an EAQEC code with this property.
If we move the vertical line and put it between the fourth and fifth qubits in (\ref{eq:Bowen's_code}), we obtain the stabilizer generators of the $[[4,1,3;1]]_{AB}$ EAQEC code in \cite{BDM06}.
These observations inspire us to find EAQEC codes that are equivalent to standard stabilizer codes and the answer is straightforward as follows.
Given a set of stabilizer generators  $\{g_1, g_2,\cdots, g_{n-k}\}$ of an $[[n,k,d]]$ stabilizer code, after Gaussian elimination (and reordering qubits as necessary), they can be written as
$g_1' \otimes Z_1$,
$h_1' \otimes X_1$,
$g_2' \otimes Z_2$,
$h_2' \otimes X_2$,
$\cdots$,
$g_c' \otimes Z_c$,
$h_c' \otimes X_c$,
$g_{c+1}' \otimes I$,
$\cdots$,
$g_{n-k-c}' \otimes I $
for some $c\leq \lfloor \frac{n-k}{2}\rfloor$ such that the  \emph{simplified generators} $g_i'$ and $h_i'$ anti-commute with each other, and they commute with other {simplified generators} $g_j', h_l'$ for $i=1, \cdots, c$ and $j,l \neq i$;
while $g_{c+1}', \cdots, g_{n-k-c}'$ commute with all the simplified generators.
Consequently, $\{g_1',\cdots, g_{n-k-c}', h_1',\cdots, h_c' \}$ defines an $[[n-c,k,d;c]]$ EAQEC code.
We summarize the above result in the following theorem and provide a lower bound on $c$.
\bt \label{thm:AB_code}
Suppose $H=[H_X|H_Z]$ is the check matrix of an $[[n,k,d]]$ standard stabilizer code.
After Gaussian elimination, $H$ can be written in the following standard form:
\begin{align}
H=\left[
\begin{array}{cc|cc}
  A& I_{s\times s}   & D &   0\\
  C &  0 & B  &   I_{s\times s}\\
    E &  0 & F & 0 \\
\end{array}\right], \label{eq:standard_form_check_matrix}
\end{align}
where $0\leq s\leq n-k$.
Then there exists an $[[n-c,k,d;c]]_{AB}$ EAQEC code for some $0\leq c\leq s$
such that the code can correct $\lfloor\frac{d-1}{2}\rfloor$-qubit errors on either Alice's or Bob's qubits.
In the case of nondegenerate quantum codes,  $s$ is bounded by \[d-1\leq s \leq \lfloor \frac{n-k}{2}\rfloor.\]
\et
\begin{proof}
An $[[n-c,k,d;c]]_{AB}$ EAQEC code is defined by the check matrix $H$ with the last $c$ columns of both $H_X$ and $H_Z$  being removed.

If $d=1$, the result is trivial. Assume $d\geq 2$.
We perform Gaussian elimination on the check matrix $H$ to pair the rows in the form (\ref{eq:standard_form_check_matrix}).
When the process of Gaussian elimination stops, either all the rows of $H$ are paired or some rows are left unpaired.
In the first case,
\[
H=\left[
\begin{array}{cc|cc}
  A& I_{s\times s}   & D &   0\\
  C &  0 & B  &   I_{s\times s}\\
\end{array}\right].
\]
We have
$
2s = n-k \geq 2(d-1)
$
by the quantum singleton bound, and thus $s\geq d-1$.
In the second case, we can find a column  in $H_X$ (or $H_Z$) that has a `1' at  the $(2s+1)$-th position and $0$'s elsewhere,
and the corresponding column in $H_Z$ (or $H_X$)  is of the form \[[u \ v \  w \ 0 \cdots 0]^T, \]
where $u$ and $v$ are two binary $s$-tuples and $w$ is 0 or 1. Then $H$ is in the following form:
\[
H=\left[
\begin{array}{ccc|ccc}
  A'& 0 & \begin{array}{ccc} 1& &0\\ &\ddots & \\ 0&&1 \end{array}   & D'&  \begin{array}{c} |\\ u^T\\| \end{array} &   0\\
\hdashline
  C'& 0 & 0   & B'&  \begin{array}{c} |\\ v^T\\| \end{array} &   \begin{array}{ccc} 1& &0\\ &\ddots & \\ 0&&1 \end{array}\\
\hdashline
  E' & \begin{array}{c} 1\\ 0\\\vdots\\0 \end{array} & 0  & F'&  \begin{array}{c} w\\ 0\\\vdots\\0 \end{array}  &  0
\end{array}\right].
\]
Note that a subset of the last $(s+1)$ columns of $H_X$ and the last $(s+1)$ columns of $H_Z$ are linearly dependent.
Thus we can find an element of weight at most $(s+1)$ corresponding to these columns that is in the normalizer group of the stabilizer group. 
In the case of nondegenerate quantum codes,  
the minimum distance is the minimum weight of an element in the normalizer group of the stabilizer group.
Since the minimum distance  of the quantum code is $d$,
this implies that  
\[s\geq d-1.\]


\end{proof}
By this theorem, we can ``move" some ancilla qubits to Bob's side for any nondegenerate stabilizer codes and obtain an EAQEC code.
The case $c=d-1$ of the above theorem is also observed by Wilde and Hsieh  \cite{WH10} from the viewpoint of purification and tracing qubits \cite{Pre99}.
Note that Shor's $[[9,1,3]]$ degenerate code \cite{Shor95} and Bacon's $[[n^2,1,n ]]$ degenerate codes \cite{Bacon06.PhysRevA.73.012340} also satisfy this theorem (by pairing up $S_i^X$ with $S_i^Z$).
It is conjectured that all degenerate codes satisfy the lower bound in Theorem \ref{thm:AB_code}.
We have checked that all the optimal quantum codes in Grassl's table \cite{Grassl} satisfy the lower bound in this theorem.

EAQEC codes that are equivalent to standard stabilizer codes can correct errors on the qubits of  both Alice and Bob.
The  decoder of the corresponding standard stabilizer code can be adopted to decode these EAQEC codes.
These codes may perform better in practice than their corresponding standard stabilizer codes, for there are fewer physical qubits transmitted through the noisy channel,
and the storage error rate is generally lower than the noisy channel error rate.



In the case of CSS codes, the standard form of a parity check matrix is
\begin{align*}
H=\left[
\begin{array}{cc|cc}
   A & I_{s\times s} &  0 & 0\\
    0 & 0 & B& I_{(n-k-s)\times (n-k-s)} \\
\end{array}\right].
\end{align*}
Consequently, we have the following theorem.
\bt
An $[[n,k,d]]$ CSS code, obtained from an  $[n,k',d]$ classical dual-containing code with $k=2k'-n$, gives
$[[n-c,k,d;c]]_{AB}$ EAQEC codes for $0\leq c\leq \frac{n-k}{2}=n-k'$.
\et
\noindent This shows that any CSS codes can be transformed into EAQEC codes that correct errors on both Alice's and Bob's qubits.
The decoding method in this scheme is exactly the same as that of standard CSS codes.
Since many stabilizer codes are based on the CSS construction,
we can take advantage of these codes while also having the power of entanglement.

\be
From  the $[[7,1,3]]$  Steane code, we obtain  a $[[4,1,3;3]]_{AB}$, a $[[5,1,3;2]]_{AB}$, and a $[[6,1,3;1]]_{AB}$  EAQEC code.
The $[[4,1,3;3]]_{AB}$ and the $[[6,1,3;1]]_{AB}$ EAQEC codes were also found by Wilde and Fattal \cite{WF10} and Shaw et al \cite{SWOKL08}, respectively:
\[
\begin{array}{cccc|ccc}
X&X&I&X&X&I&I\\
X&X&X&I&I&X&I\\
X&I&X&X&I&I&X\\
Z&Z&I&Z&Z&I&I\\
Z&Z&Z&I&I&Z&I\\
Z&I&Z&Z&I&I&Z
\end{array}.
\]
\ee

The quantum singleton bound says that for an $[[n,k,d]]$ quantum code, \[n-k\geq 2(d-1).\]
If the parameters $n,k,d$ of a standard stabilizer code achieve the quantum singleton bound, or
$n-k=2(d-1),$
then EAQEC codes equivalent to these standard stabilizer codes will achieve the singleton bound for EAQEC codes: \[n+c-k\leq 2(d-1),\] and we have the following theorem.
\bt
A standard stabilizer code that achieves the quantum singleton bound gives an
$[[n-c,k,d;c]]_{AB}$ EAQEC code for some $c$ that achieves the singleton bound for EAQEC codes.
\et

\be
The $[[3,1,3;2]]_{AB}$ and $[[4,1,3;1]]_{AB}$ EAQEC codes derived from the five qubit code achieve the singleton bound for EAQEC codes.
\ee

\be
Using MAGMA \cite{MAGMA} to find optimal standard stabilizer codes,
we obtain several optimal EAQEC codes that achieve the linear programming bounds in \cite{LBW10} by Theorem \ref{thm:AB_code}: 
\begin{align*}
&[[15,10,4;5]]_{AB},
[[14,11,3;3]]_{AB},
[[13,9,4;4]]_{AB},\\
&[[13,10,3;3]]_{AB},
[[12,9,3;3]]_{AB},
[[11,8,3;3]]_{AB},\\
&[[10,6,4;4]]_{AB},
[[10,7,3;3]]_{AB},
[[9,6,3;3]]_{AB},\\
&[[7,4,3;3]]_{AB},
[[8,4,4;4]]_{AB},
[[6,2,4;4]]_{AB},\\
&[[7,3,3;1]]_{AB},
[[6,3,3;2]]_{AB},
[[6,1,5;5]]_{AB},\\
&[[4,1,3;1]]_{AB},
[[4,1,3;3]]_{AB},
[[3,1,3;2]]_{AB}.
\end{align*}
We also found the following EAQEC codes that have the highest minimum distance for fixed $n$ and $k$ to the best of our knowledge:
\begin{align*}
&[[15,4,8;11]]_{AB},
[[13,5,6;8]]_{AB},
[[12,6,5;6]]_{AB}.
\end{align*}

\ee

\section{Quantum Codes with Two Encoders} \label{sec:two_encoder}

In the previous section, we discussed EAQEC codes that are equivalent to standard stabilizer codes.
  Most optimal EAQEC codes are not equivalent to any standard stabilizer codes, such as the entanglement-assisted repetition codes \cite{LB10,LBW10}.
We would like to exploit the high error-correcting ability of these quantum codes even in the presence of storage errors on Bob's side.
This can be achieved by using another quantum code to protect Bob's qubits.

Assume Alice uses the encoding operator $U_A$ of an $[[n,k,d_A;c]]$ EAQEC code to protect her information qubits.
Suppose also that there are $m-c>0$ ancilla qubits on Bob's side, and that Bob applies the encoding operator $U_B$ of an $[[m,c,d_B]]$ standard stabilizer code to protect his $c$ qubits.
The encoding operator on the whole is $U_A\otimes U_B$.
We use the notation $[[n,k,d_A;c]]+[[m,c,d_B]]$ to represent such a composite quantum code.
If there are no ancillas on Bob's side, the set of stabilizer generators is equivalent to that of an EAQEC code for some encoding operator $U_A'\otimes I^B$ after Gaussian elimination.

The initial state is
$
\ket{\psi}\otimes\ket{0}^{\otimes n-c-k}\otimes \ket{\Phi_+}^{\otimes c}\otimes \ket{0}^{\otimes m-c}.
$
The encoded state has the following stabilizer generators:
$g_1 \otimes  u_1$,
$h_1 \otimes v_1$,
$g_2 \otimes u_2$,
$h_2 \otimes v_2$,
$\cdots$,
$g_c \otimes u_c$,
$h_c \otimes v_c$,
$g_{c+1} \otimes I$,
$\cdots$,
$g_{n-k-c} \otimes I$,
$I \otimes   u_{c+1}$,
$\cdots$,
$I\otimes u_m$,
where $U_B Z_i^B U_B^{\dag} =u_i$ and $U_B X_j^B U_B^{\dag} =v_j$.

A straightforward decoding process of the  $[[n,k,d_A;c]]+[[m,c,d_B]]$ quantum code is that
Bob first decodes his $c$ ebits in the $[[m,c,d_B]]$ quantum code, and then he decodes the $k$ information qubits hiding in the  $[[n,k,d_A;c]]$ EAQEC code.
Or we can treat the combination code as an $[[n+m,k,d_c]]$ code, which has a more complicated decoding circuit but a potentially higher error-correcting ability.

Under what condition is an $[[n,k,d;c]]$ EAQEC code not equivalent to an $[[n+c,k,d]]$ standard stabilizer code?
An $[[n,k,d;c]]$ EAQEC code satisfies the Hamming bound for EAQEC codes \cite{Bowen02}:
\begin{align*}
\sum_{j=0}^t 3^j {n\choose j}\leq 2^{n-k+c}.
\end{align*}
for $t = {\lfloor \frac{d-1}{2}\rfloor}$.  If the parameters do not satisfy the quantum Hamming bound for (nondegenerate) quantum codes:
\begin{align*}
\sum_{j=0}^t  3^j {n+c \choose j}\leq 2^{n-k+c},
\end{align*}
an $[[n+c,k,d]]$ code does not exist.
\be
Consider the $[[7,1,5;2]]$ code in \cite{LB10}. We have
$
\sum_{j=0}^t 3^j {n\choose j}= 211 < 2^{n-k+c}=256.
$
However,
$
\sum_{j=0}^t 3^j {n+c\choose j}= 352 > 2^{n+c-k}=256.
$
Hence  there is no $[[9,1,5]]$ code.
\ee

However, the Hamming bound is not tight. For example, the $[[7,1,5;3]]$ code is not equivalent to any standard code,
but the parameters $[[10,1,5]]$ satisfy the quantum Hamming bound.
A better bound, such as the linear programming bound, can be applied here.
If the parameters $n+c, k, d$  violate any upper bound on the minimum distance of the quantum code,
then such an $[[n+c,k,d]]$ standard stabilizer code does not exist, and the $[[n,k,d;c]]$ EAQEC code is not equivalent to any standard stabilizer code.
We can check the result for $n+c\leq 30$ from the tables of stabilizer codes in \cite{CRSS98} and \cite{Grassl}.

Several EAQEC codes that not equivalent to standard codes were found in \cite{LB10}:
$[[n,1,n;n-1]]$ for $n$ odd, $[[n,1,n-1;n-1]]$ for $n$ even,
$[[7,1,5;2]]$, $[[7,1,5;3]]$, $[[7,2,5;5]]$, $[[9,1,7;4]]$, $[[9,1,7;5]]$, $[[9,1,7;6]]$, $[[9,1,7;7]]$, $[[8, 2, 5; 4]]$, $[[8, 3, 5; 5]]$, $[[13, 3, 9; 10]]$,
 $[[13, 1, 11; 11]]$, $[[13, 1, 11; 10]]$, $[[13, 1, 9; 9]]$, $[[13, 1, 9; 8]]$.
Also, the $[[15,7,6;8]]$, $[[15,8,6;7]]$, $[[15,9,5;6]]$ EAQEC codes are not equivalent to any known standard quantum stabilizer code.

\be
The $[[n,1,n;n-1]]$ EAQEC codes for $n$ odd saturate the quantum Hamming bound with equality as follows.
The number of correctable $X$ or $Z$ errors is $\left\lfloor  \frac{n-1}{2}\right\rfloor$ and the number of correctable error syndromes is
\[
\left({n\choose 0}+{n\choose 1}+ \cdots +{n\choose \lfloor \frac{n-1}{2}\rfloor}\right)^2=2^{2(n-1)}.
\]
Suppose Alice uses the $[[5,1,5;4]]$ entanglement-assisted repetition code and Bob applies the optimal $[[10,4,3]]$ quantum code to protect his $4$ ebits.
Then the whole quantum code $[[5,1,5;4]]+[[10,4,3]]$ can protect two channel noise errors on the $5$ qubits that Alice sends through the channel, and one storage error on Bob's $10$.
On the other hand, the optimal quantum code using $15$ qubits to encode one information qubit is the $[[15,1,5]]$ quantum code \cite{Grassl}, and it can correct an arbitrary two qubit error.
\ee

Compared to an $[[n+m,k,d]]$ standard stabilizer code, the number of qubits going through the noisy channel is much less  using the  $[[n,k,d_A;c]]+[[m,c,d_B]]$ quantum code.
Consequently, as long as rate the storage error is reasonably small compared to the channel error rate,
the  $[[n,k,d_A;c]]+[[m,c,d_B]]$ quantum code has better efficiency, while keeping  the same error-correcting ability against the channel noise.
The performances of different coding schemes will be compared in Section \ref{sec:comparison}.

\begin{figure}
\begin{center}
\includegraphics[height=3cm]{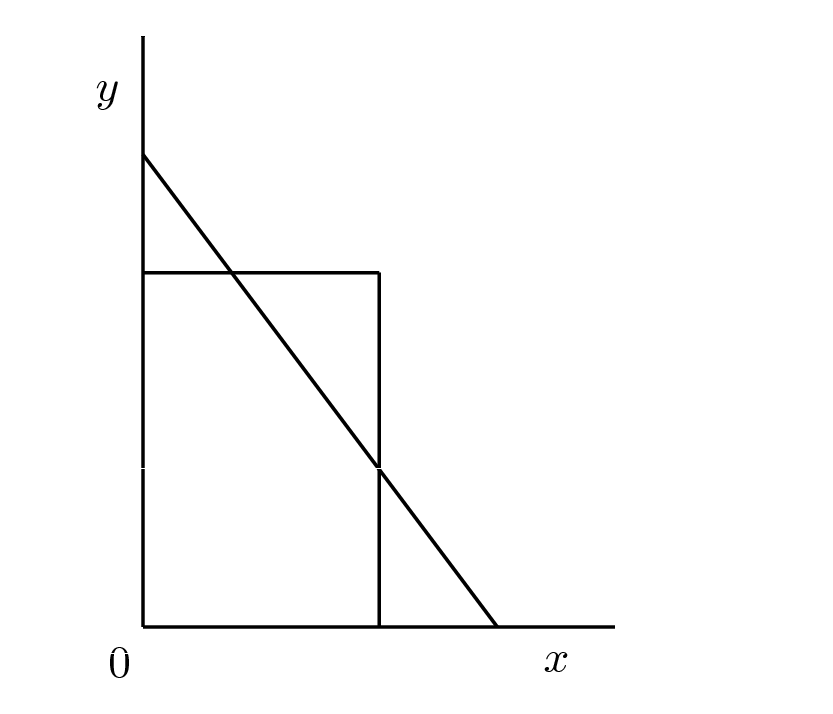}
{\caption{Illustration of the weight distributions of the syndrome representatives of the two decoding method.
} \label{fig:weight_distribution_syndrome_representatives}}
\end{center}
\end{figure}

For combination EAQEC codes, the syndrome representatives can be chosen as in the case of standard stabilizer codes if we treat the code as a single stabilizer code.
If we use two sequential decoders, we choose two sets of syndrome representatives $T_A$, $T_B$
as in the case of standard stabilizer codes.
Observe that $T=T_A\times T_B$ is the set of syndrome representatives of the combination code.
The weight distributions of the syndrome representatives of the two decoding method are illustrated in Fig. \ref{fig:weight_distribution_syndrome_representatives}.
The $x$-axis and $y$-axis represent the weights on Bob's and Alice's qubits, respectively.
The weight distribution of the syndrome representatives using two sequential decoders is always a rectangle.
If the error probabilities are the same on  Alice's and Bob's qubits,
the weight distribution of the syndrome representatives using a single decoder looks like a triangle.
If the error probabilities are different on Alice's and Bob's qubits, the shape varies according to the error probabilities.
Note that the area of the triangle is equal to the area of the rectangle.


\section{Channel Fidelity} \label{sec:channel_fidelity}

%

\subsection{Formula for the Channel Fidelity over the Depolarizing Channel}
\label{subsec:fc_formula}

Let $\mathcal{E}$ be a quantum channel operating on the input state $\rho$, which lies in a state space $\mathcal{H}'$ of dimension $m$,
and the output state $\mathcal{E}(\rho)$ also lies in $\mathcal{H}'$.
Suppose the quantum channel $\mathcal{E}$ has the operator-sum representation $\mathcal{E}(\rho)=\sum_{i}E_i \rho E_i^{\dag}$,
where the operation elements $\{ E_i\}$ satisfy $\sum_i E_i^{\dag}E_i =I$.
Let $\ket{\psi}= \frac{1}{m^{1/2}} \sum_{j} \ket{j}\ket{j}$, where $\{ \ket{j}\}$ is a basis of the input space $\mathcal{H}'$.
 The channel fidelity  \cite{Schumacher96,RW06} of  $\mathcal{E}$  is defined as
 \begin{align*}
F_C(\mathcal{E}) =&\bra{\psi}\left(I\otimes \mathcal{E}\ket{\psi}\bra{\psi} \right)\ket{\psi}
= \frac{1}{m^{2}} \sum_{i} |\tr(E_i)|^2.
 \end{align*}

The channel fidelity can be used as a measure of the performance of a quantum error-correcting code over a noisy channel.
Suppose the $k$-qubit information state $\ket{\psi}$, after being encoded by the encoding channel $\mathcal{U}: \mathcal{H}^{\otimes k}\rightarrow \mathcal{H}^{\otimes n}$, is transmitted through a noisy channel $\mathcal{N}: \mathcal{H}^{\otimes n}\rightarrow \mathcal{H}^{\otimes n}$,
and then decoded by the decoding channel $\mathcal{D}: \mathcal{H}^{\otimes n}\rightarrow \mathcal{H}^{\otimes k}$.
Then $F_C(\mathcal{D}\mathcal{N}\mathcal{U})$ serves  as a merit function of the quantum code with encoding-decoding pair $\{\mathcal{U},\mathcal{D}\}$ over the noisy channel $\mathcal{N}$.

The encoding channel $\mathcal{U}$ can be written as
\[
    \mathcal{U}(\ket{\psi}\bra{\psi})= U_E(I^{\otimes k} \otimes \ket{0}^{\otimes n-k}) \ket{\psi}\bra{\psi} (I^{\otimes k} \otimes \bra{0}^{\otimes n-k})U_E^{\dag},
\]
for some unitary Clifford encoder $U_E$.
We assume there are $(n-k)$ ancillas and we implicitly use the equation
$$ \left(I^{\otimes k} \otimes \ket{0}^{\otimes n-k}\right) \ket{\psi}= \ket{\psi}\otimes \ket{0}^{\otimes n-k}$$
for any $k$-qubit state $\ket{\psi}$.
The decoding channel $\mathcal{D}$ consists of the following steps: syndrome measurement $\{P_s\}$, correction $\{C_s\}$, decoding $U^{\dag}_E$, and partial trace of the ancillas,
which have been explained in Section \ref{sec:basics}.
The overall process of the decoding channel is
\begin{align*}
    \mathcal{D}(\rho')=& \tr_A \left(\sum_s  U_E^{\dag }C_s P_s \rho' P_s C_r^{\dag}U_E\right)\\
    =& \sum_{l}\sum_s  (I^{\otimes k}\otimes \bra{l} )U_E^{\dag }C_s P_s \rho' P_s C_r^{\dag}U_E (I^{\otimes k}\otimes \ket{l} ),
\end{align*}
where $\{\ket{l}\}$ is the standard basis for the state space of the  $(n-k)$ ancilla qubits, and $\ket{l}$ has a binary representation $\ket{l_1 l_2\cdots l_{n-k}}$.
Thus $\mathcal{D}$ has operation elements $\{ (I^{\otimes k}\otimes \bra{l} )U_E^{\dag }C_s P_s  \}_{l,s}$.
Suppose the noisy channel $\mathcal{N}$ is the independent $n$-fold depolarizing channel $T_p^{\otimes n}$ with
operation elements $\{\sqrt{p_i} E_i\}$, where $p_i$ is defined in (\ref{eq:probability_def}).

The composite channel $\mathcal{D}\mathcal{N}\mathcal{U}$ has
operation elements
$$\left\{ W_{l,i,s}\triangleq(I^{\otimes k}\otimes \bra{l} )U_E^{\dag }C_s P_s \sqrt{p_i}E_i  U_E(I^{\otimes k} \otimes \ket{0}^{\otimes n-k})  \right\}_{l,i,s}$$
and its channel fidelity is
\begin{align*}
F_C( \mathcal{D}\mathcal{N}\mathcal{U})=& \frac{1}{2^{2k}}\sum_{l,s,i}\left|W_{l,i,s} \right|^2.
\end{align*}
There are a total of $2^{n-k}\cdot 2^{n-k}\cdot 4^n =4 ^{2n-k}$ terms in the sum, and each term is a product of one $(2^k\times 2^{n})$ matrix, five $(2^n\times 2^n)$ matrices, and one $(2^n\times 2^k)$ matrix.
Thus the complexity of the calculation of the channel fidelity is $\Omega(n\cdot 4^{3n-2k})$ (the complexity of multiplication of $(2^n\times 2^n)$ matrices is $\Omega(n2^{n})$), which is almost impossible to calculate for $n>10$.
We will show how to reduce the complexity to something more manageable.

\bl \label{lemma:measurement_syndrome}
If the error syndrome of $E_i$ is $s_i$, then
\[\left| \tr\left(   W_{l,i,s}\right) \right|^2=0\] for $s\neq s_i$.
In addition, the measurement result is $s_i$ with certainty.
\el
It is straightforward to check that $P_s E_i U_E(I^{\otimes k} \otimes \ket{0}^{\otimes n-k})=0$ from the facts that $P_s$'s are orthogonal to each other and $P_0$ is the projector on the code space,
and the above lemma follows naturally.
%
Next we show that only the error operators in $T\times \mathcal{S}$ have nonzero contribution  to the channel fidelity. 
\bl \label{lemma:correctable_error_tofidelity}
\begin{align*}
\frac{1}{2^{2k}}\sum_l \left| \tr\left( W_{l,i,s_i}\right) \right|^2
=&
\left\{%
\begin{array}{ll}
    p_i, & \hbox{ if $E_i\in  T\times \mathcal{S}$;} \\
    0, & \hbox{otherwise.} \\
\end{array}%
\right.
\end{align*}

\el
\begin{proof}
It is straightforward to verify the lemma for the case that $E_i \in  T\times \mathcal{S}$. 
Now assume $E_i\notin T\times \mathcal{S}$ and we have $C_{s_i}E_i\in \mathcal{N}(\mathcal{S})\backslash \mathcal{S}$.
Since $U_E$ is a Clifford unitary operator, $M_1\otimes M_2\triangleq  U_E^{\dag}C_{s_i}E_i U_E\in \mathcal{G}_n$ is a Pauli operator, where
$M_1\in \mathcal{G}_k$ and $M_2\in \mathcal{G}_{n-k}$.
Since $E_i\notin T\times \mathcal{S}$, we have  $M_1$ not equal to the identity and $\tr(M_1)=0$.
Let the matrix representations of $M_1$ and $M_2$ be $[a_{i,j}]$ and  $[b_{i,j}]$, respectively.
Then
\begin{align*}
&\sum_l  \left| \tr\left((I^{\otimes k}\otimes \bra{l} )U_E^{\dag }C_{s_i} P_{s_i} E_i  U_E(I^{\otimes k} \otimes \ket{0}^{\otimes n-k})\right) \right|^2\\
=&\sum_l  \left| \tr\left((I^{\otimes k}\otimes \bra{l})M_1\otimes M_2 (I^{\otimes k} \otimes \ket{0}^{\otimes n-k})\right) \right|^2\\
=& \sum_l  \left| b_{l,1}\cdot \tr(M_1) \right|^2=0,
\end{align*}
where the second equality follows by explicitly writing down the matrix multiplications of $(I^{\otimes k}\otimes \bra{l})M_1\otimes M_2 (I^{\otimes k} \otimes \ket{0}^{\otimes n-k})$. 

\end{proof}

We derive the following theorem with the help of  Lemma \ref{lemma:measurement_syndrome} and  Lemma \ref{lemma:correctable_error_tofidelity}.
\bt \label{thm:FC formula}
The channel fidelity of a quantum code, with a stabilizer group $\mathcal{S}$ and a set of syndrome representatives $T$,  over the depolarizing channel, is
the weight enumerator of  the probability distribution $\{q_w\}$ of the elements in $T\times \mathcal{S}$.
\et
\begin{proof}
\begin{align}
F_C( \mathcal{D}\mathcal{N}\mathcal{U}) \nonumber
=& \frac{1}{2^{2k}}\sum_{l,i} \left| \tr\left( W_{l,i,s_i}\right) \right|^2 \nonumber\\
=& \sum_{E_i\in T\times \mathcal{S}}p_i  \nonumber\\
=& \sum_{w=0}^n  a_w q_w, \label{eq:FC_formula}
\end{align}
where $a_w$ is the number of $E_i \in T\times \mathcal{S}$ of weight $w$, and $q_w$ is defined in (\ref{eq:probability_def}).
The first equality follows by Lemma \ref{lemma:measurement_syndrome} and the second equality follows by Lemma \ref{lemma:correctable_error_tofidelity}.

\end{proof}
There are $4^{n-k}$ terms in (\ref{eq:FC_formula}) and each term is generated by vector addition.
The complexity is now reduced to $\Omega(n\cdot 4^{n-k})$.
From the above theorem, we find that the channel fidelity for the depolarizing channel is the probability of correctable errors.
For a single information-qubit code, the channel fidelity is the probability that the information qubit can be  correctly recovered.
The formula for the channel fidelity of several quantum codes is shown in Table. \ref{table:FC_formula}.

The channel fidelity of the quantum code defined by a stabilizer group $\mathcal{S}$ over a depolarizing channel
 depends on the syndrome representatives $T$ and the stabilizer group $\mathcal{S}$,
 and it can be optimized over the choices of $T$.
(This is to optimize over different decoding schemes.)



\begin{table}

\begin{center}
\begin{tabular}{|c|c|} %
\hline
  Type of codes & Channel fidelity \\
   \hline
  bit-flip code & $1- \frac{3}{2}p+\frac{9}{8}p^2-\frac{3}{8}p^3$\\
\hline
$[[4,2,2]]$  code  & $1-\frac{3}{2}p+\frac{3}{4}p^2 $\\
\hline
$[[5,1,3]]$ code  & $1-\frac{45}{8}p^2+\frac{75}{8}p^3-\frac{45}{8}p^4+\frac{9}{8}p^5 $\\
\hline
$[[7,1,3]]$  code  & $1-\frac{147}{16}p^2+\frac{189}{8}p^3-\frac{1785}{64}p^4+\frac{1155}{64}p^5 $\\
   & $-\frac{399}{64}p^6+\frac{57}{64}p^7$\\
\hline
$[[8,3,3]]$ code & $1-\frac{245}{16}p^2+ \frac{1449}{32}p^3-\frac{8029}{128}p^4 +\frac{12743}{256}p^5$\\
&$-\frac{2961}{128}p^6+\frac{763}{128}p^7-\frac{21}{32}p^8$\\
\hline
$[[9,1,3]]$  code  & $1-9p^2+\frac{195}{8}p^3-\frac{1071}{32}p^4+\frac{945}{32}p^5 $\\
   & $-\frac{567}{32}p^6+\frac{225}{32}p^7-\frac{27}{16}p^8+\frac{1545}{8192}p^9$\\
\hline
\end{tabular}
{\caption{channel fidelity of different quantum codes in  the depolarizing channel}
 \label{table:FC_formula}}
\end{center}
\end{table}

\subsection{Channel Fidelity for EAQEC Codes}\label{subsec:fc_formula_EAQEC}
Suppose that Alice uses a noisy channel $\mathcal{N}_A$ to communicate with Bob and Bob's ebits suffer from a storage error channel $\mathcal{N}_B$.
\bd The channel fidelity  \cite{Schumacher96,RW06} of an EAQEC code with encoding and decoding processes $\{\mathcal{U}, \mathcal{D} \}$  over the noisy channel $\mathcal{N}_A\otimes \mathcal{N}_B$ is
\[
F_C(\mathcal{D} (\mathcal{N}_A\otimes \mathcal{N}_B) \mathcal{U} ).
\]
\ed
Suppose an $[[n,k,d;c]]$ EAQEC code is defined by a simplified stabilizer group $\mathcal{S}'=\langle g_1', \cdots, g_{n-k}',$ $h_1',\cdots, h_c' \rangle \in \mathcal{G}_n$.
We extend the simplified stabilizer group to a stabilizer group $\mathcal{S}=\langle g_1,\cdots, g_{n-k},$ $h_1,\cdots, h_c \rangle \in \mathcal{G}_{n+c}$,
where $g_i= g_i'\otimes Z_{i}$, $h_i= h_i'\otimes X_{i}$, for $i=1,\cdots,c$,   and $g_j= g_j'\otimes I$ for $j=c+1,\cdots, n-k$.
We choose a set $T$ of syndrome representatives for the noisy channel $\mathcal{N}_A \otimes \mathcal{N}_B$.
Let $p_a$ and $p_b$ be the depolarizing rate of  $\mathcal{N}_A$ and $\mathcal{N}_B$, respectively.
Now we can apply Theorem \ref{thm:FC formula} to find a formula for the channel fidelity of EAQEC codes.
\bt \label{thm:FC formula_EAQECC}
The channel fidelity of an EAQEC code with a  stabilizer group $\mathcal{S}$ and the set of syndrome representatives $T$
over the depolarizing channel $\mathcal{N}_A\otimes \mathcal{N}_B$ is the weight enumerator of the probability distribution  $\{q_wr_{w'}\}$ of the elements in $T\times \mathcal{S}$,
where $p_w$ and $r_{w'}$ are defined in (\ref{eq:probability_def_Na}) and (\ref{eq:probability_def_Nb}).
\et
Note that the channel fidelity can be optimized over the choice of $T$.
Since the two noisy channels $\mathcal{N}_A$ and $\mathcal{N}_B$ have different error rates, error operators on these two channels should be differently weighted.
In the extreme case of $p_b=0$  and $r_0=1$, the error-correction condition for EAQEC codes says that
$\{E_i\}$ is a set of correctable error operators if $E_i^{\dag} E_j \notin \mathcal{N}(\mathcal{S'})\backslash \mathcal{S}_I$,
where $\mathcal{S}_I$ is the isotropic subgroup generated by $\{ g_{c+1}', \cdots, g_{n-k}'\}$.
Note that $r_0=1$, and $r_i=0$ for $i\neq 0$. We can similarly formulate the channel fidelity.
\bc
The channel fidelity of an EAQEC code with a  stabilizer group $\mathcal{S}$ and a set of syndrome representatives $T$
over the depolarizing channel $\mathcal{N}_A\otimes I_B$ is the weight enumerator of  $T\times \mathcal{S}_I$, which is a polynomial in $\{q_w\}$.
\ec
Therefore, we would like to choose $T$ to consist of likely error operators from the noisy channel $\mathcal{N}_A$.

\begin{figure}
[ptb]
\begin{center}
\includegraphics[
height=4.5cm,
]%
{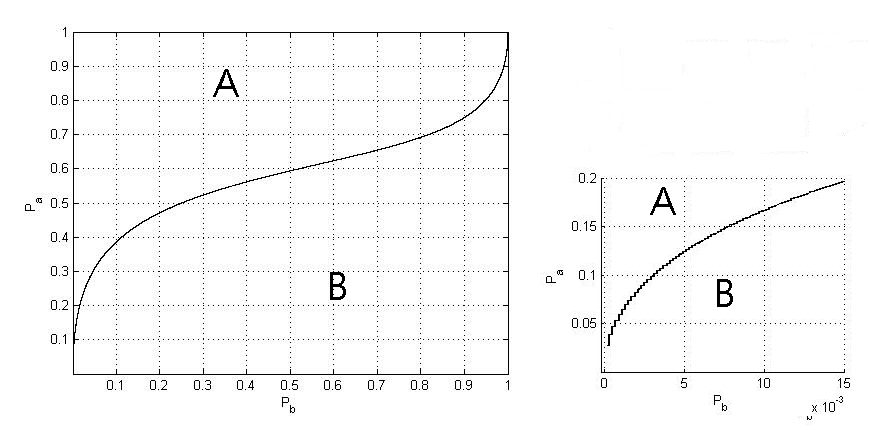}
{\caption{ Comparison of two $[[3,1,3;2]]$ EAQEC codes in terms of channel fidelity. The
$[[3,1,3;2]]_{AB}$ code performs better in Region B, while the
repetition code performs better in Region A. The region for $p_a<0.015$ and $p_b<0.2$ is enlarged on the right.}
\label{fig:EA-3qubit_code}}
\end{center}
\end{figure}
\begin{figure}
[ptb]
\begin{center}
\includegraphics[
height=6.5cm,
]%
{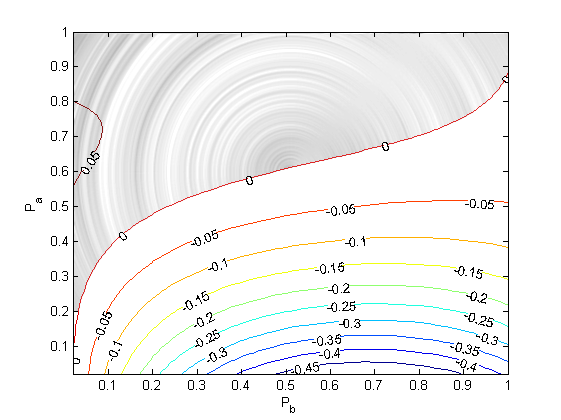}
{\caption{(Color online) Contour plot of the difference between the two fidelities: $F_C(\mathcal{D}_3^{re}(\mathcal{T}_{p_a}^{\otimes 3}\otimes \mathcal{T}_{p_b}^{\otimes 2}) \mathcal{U}_3^{re})-F_C(\mathcal{D}_3^{AB}(\mathcal{T}_{p_a}^{\otimes 3}\otimes \mathcal{T}_{p_b}^{\otimes 2}) \mathcal{U}_3^{AB})$.
The shaded region is where $F_C(\mathcal{D}_3^{re}(\mathcal{T}_{p_a}^{\otimes 3}\otimes \mathcal{T}_{p_b}^{\otimes 2}) \mathcal{U}_3^{re})-F_C(\mathcal{D}_3^{AB}(\mathcal{T}_{p_a}^{\otimes 3}\otimes \mathcal{T}_{p_b}^{\otimes 2}) \mathcal{U}_3^{AB})>0$.}
\label{fig:EA-3qubit_code3D}}
\end{center}
\end{figure}
\be
We compare the channel fidelities of the $[[3,1,3;2]]$ entanglement-assisted (EA) repetition code \cite{LB10}, $F_C(\mathcal{D}_3^{re}(\mathcal{T}_{p_a}^{\otimes 3}\otimes \mathcal{T}_{p_b}^{\otimes 2}) \mathcal{U}_3^{re})$, and
Bowen's $[[3,1,3;2]]_{AB}$ quantum code \cite{Bowen02}, $F_C(\mathcal{D}_3^{AB}(\mathcal{T}_{p_a}^{\otimes 3}\otimes \mathcal{T}_{p_b}^{\otimes 2}) \mathcal{U}_3^{AB})$.
Bowen's $[[3,1,3;2]]_{AB}$ quantum code is equivalent to the five qubit code and has $T=\{I,$ $X_1,\cdots,X_5,$ $Z_1,\cdots,Z_5,$ $Y_1,\cdots,Y_5 \}$,
while the $[[3,1,3;2]]$ EA repetition code is designed under the assumption that Bob's qubits are perfect and has $T=\{I, X_1,\cdots,X_3,$ $Z_1,\cdots,Z_3,$ $Y_1,\cdots,Y_3,$ $X_1Z_2,X_1Z_3,$ $Z_1X_2,X_2Z_3,$ $Z_1X_3,Z_2X_3 \}$.
We have
\begin{align*}
&F_C(\mathcal{D}_3^{AB}(\mathcal{T}_{p_a}^{\otimes 3}\otimes \mathcal{T}_{p_b}^{\otimes 2})
\mathcal{U}_3^{AB})= q_0r_0+
9q_1r_0+6q_3r_0+6q_0r_1\\
&+36q_2r_1+54q_3r_1+18q_1r_2+81q_2r_2+45q_3r_2,
\end{align*}
and
\begin{align*}
&F_C(\mathcal{D}_3^{re}(\mathcal{T}_{p_a}^{\otimes 3}\otimes \mathcal{T}_{p_b}^{\otimes 2})
\mathcal{U}_3^{re})= q_0r_0+ 9q_1r_0+6q_2r_0\\
&+18q_1r_1+38q_2r_1+40q_3r_1+18q_1r_2+55q_2r_2+71q_3r_2.
\end{align*}
The channel fidelities of these two EAQEC codes are compared in Fig. \ref{fig:EA-3qubit_code} and Fig. \ref{fig:EA-3qubit_code3D}.
In Fig. \ref{fig:EA-3qubit_code}, the curve of the boundary between the two regions passes the origin.
The region A in Fig. \ref{fig:EA-3qubit_code} corresponds to the shaded part in Fig. \ref{fig:EA-3qubit_code3D}.
In region B, Bowen's code is better than the EA repetition code.

In the extreme case of $p_b=0$, we have
\[
F_C(\mathcal{D}_3^{AB}(\mathcal{T}_{p_a}^{\otimes 3}\otimes I^{\otimes 2}) \mathcal{U}_3^{AB})=
q_0+ 9q_1+6q_3,
\]
and
\[
F_C(\mathcal{D}_3^{re}(\mathcal{T}_{p_a}^{\otimes 3}\otimes I^{\otimes 2}) \mathcal{U}_3^{re})=
q_0+ 9q_1+6q_2.
\]
The EA repetition code corrects more lower-weight errors, and hence it has higher channel fidelity.
\ee

The channel fidelity of the  $[[n,k,d_A;c]]+[[m,c,d_B]]$ quantum code depends on  the decoding process.
If we treat the combination code as an $[[n+m,k,d_C]]$ code,
its channel fidelity can be computed as Theorem \ref{thm:FC formula_EAQECC}.
If we use two sequential decoders, it's different.
First, the $4^{m}$ possible error operators on Bob's  side collapse to
 $4^c$ logical errors. Each can be obtained from $4^{m-c}$ error operators, after the decoding process of the $[[m,c,d_B]]$ quantum code.
If the $4^m$  errors occur uniformly, the decoded errors also occur uniformly.
However, this is not the case for the depolarizing channel $\mathcal{N}_B$:
the $4^c$ errors  on Bob's ebits occur according to a distribution $\{r'_{w'}\}$ that depends on the decoding process. 
We can find the channel fidelity as in Theorem \ref{thm:FC formula_EAQECC},
except that the errors on Bob's ebits follow a new distribution $\{ q_w r'_{w'}\}$.
We will compare the channel fidelities of these two decoding methods in Section \ref{sec:comparison}.


\subsection{Approximation of Channel Fidelity}{\label{subsec:monte_carlo}}
The number of terms in the formula for channel fidelity over the depolarizing channel is $4^{n+c-k}$, which grows exponentially in $n+c-k.$
It is difficult even to build a look-up table for decoding when $n$ is large.
However, it is possible to approximate  the channel fidelity efficiently.

The channel fidelity of a quantum code over a depolarizing channel is the probability that the decoder output is correct, which can be lower bounded by
 \begin{align*}
    \mbox{Pr}&( \{ \mbox{correctable errors} \} ) \\
    &\geq \mbox{Pr}( \{ \mbox{syndrome representatives} \} ),
 \end{align*}
 or
  \begin{align*}
    \mbox{Pr}&( \{ \mbox{correctable errors} \} )\\
&\geq   \mbox{Pr}( \{ \mbox{errors of weight less than or equal to $\lfloor\frac{d-1}{2}\rfloor$} \} ),
 \end{align*}
 where $d$ is the minimum distance of the quantum code.
 When the depolarizing rate is low ($<0.2$), these bounds are fairly tight and are good approximations of the true channel fidelity.
Dong et al. defined an ``infidelity" function to characterize the performance of quantum codes \cite{DDJCY09}, which is an approximation of the channel fidelity  in the case of $k=1$.

\begin{figure}
\begin{center}
\includegraphics[height=6cm]{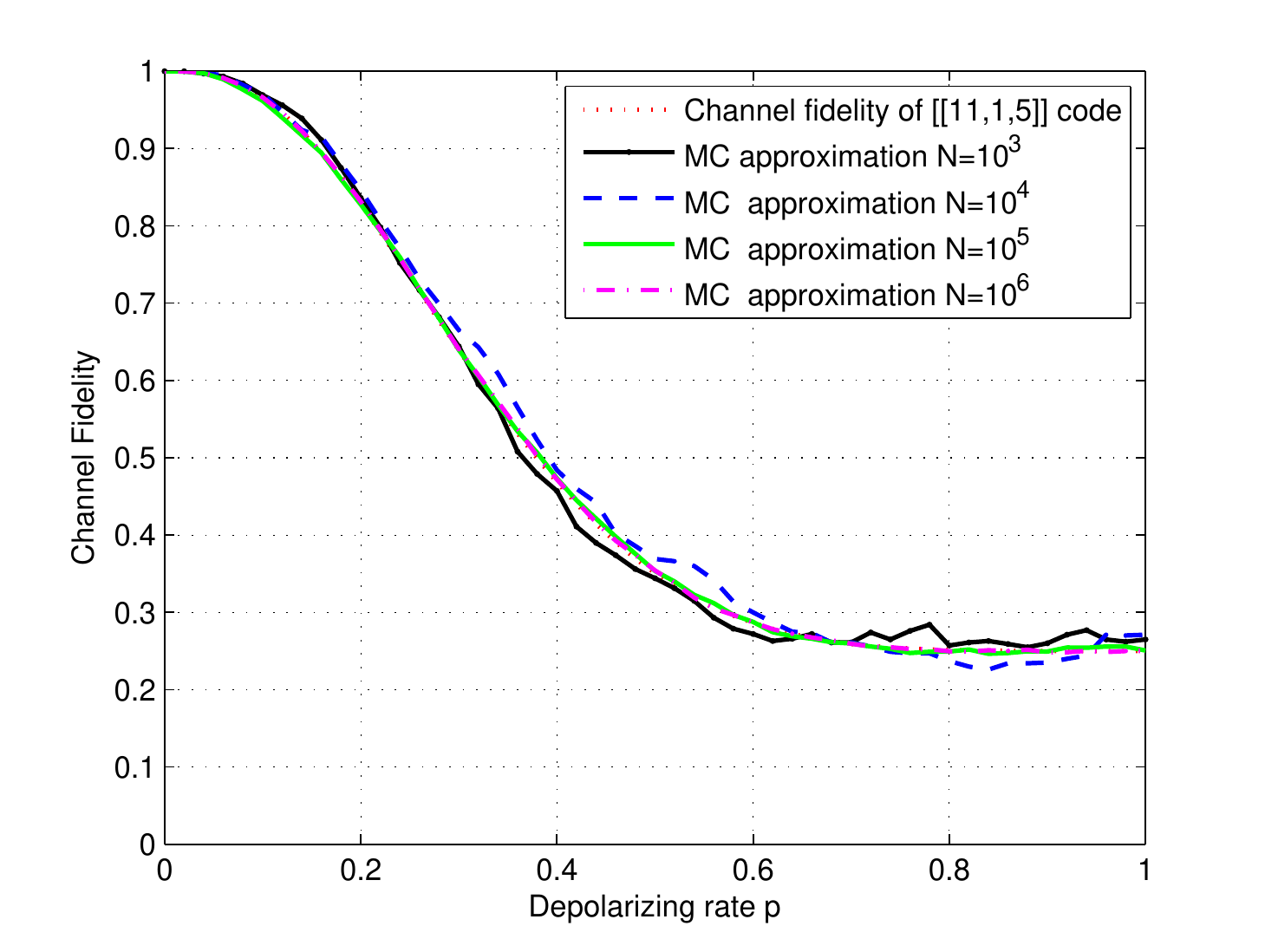}
{\caption{(Color online) The approximations of the channel fidelity of the $[[11,1,5]]$ quantum code.} \label{fig:mc_11_1_5}}
\end{center}
\end{figure}

\begin{figure}
\begin{center}
\includegraphics[height=6cm]{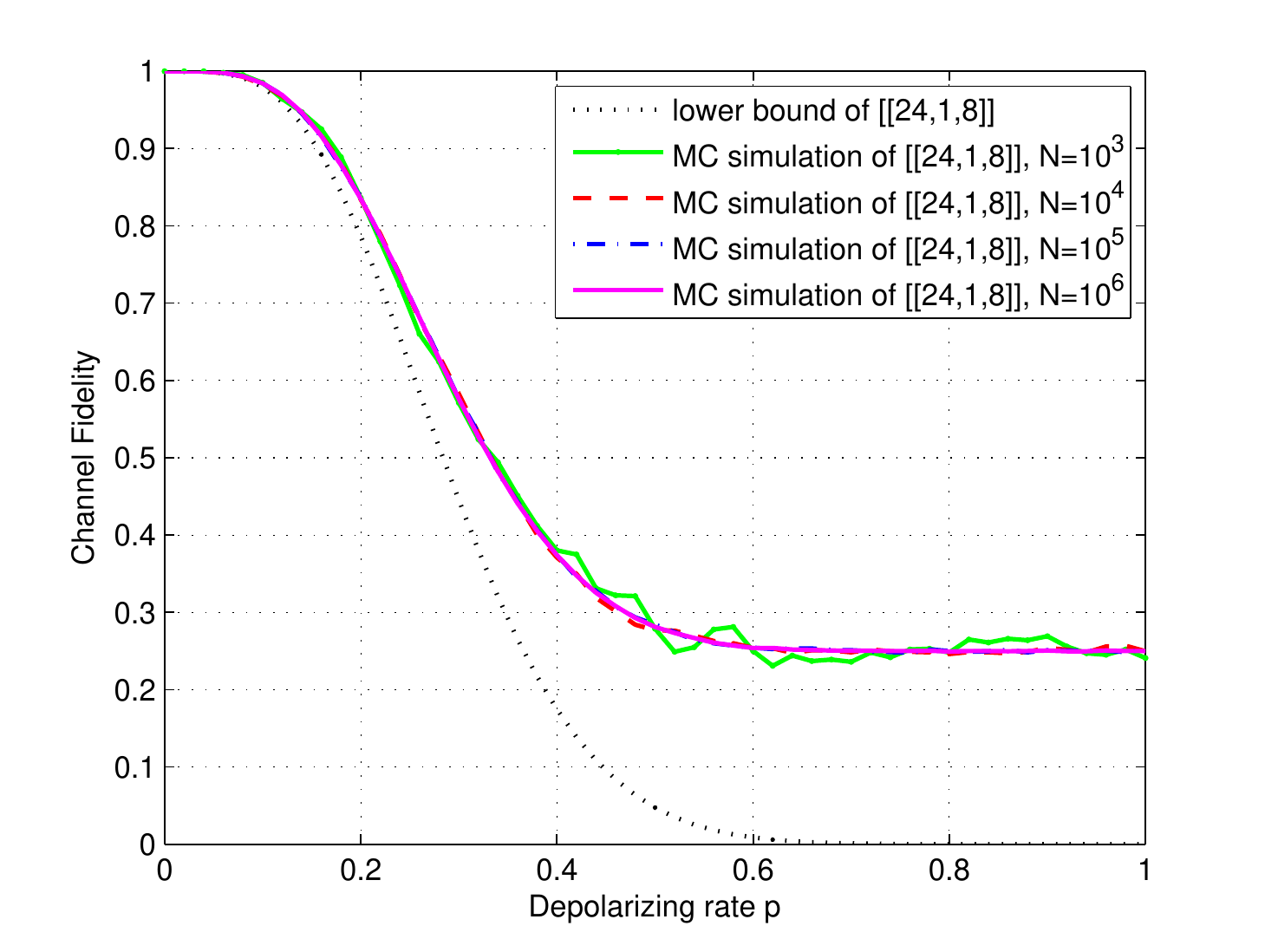}
{\caption{(Color online) The approximations of the channel fidelity of the $[[24,1,8]]$ quantum code.} \label{fig:mc_13_1_8_11}}
\end{center}
\end{figure}

We can also apply Monte Carlo methods to approximate the channel fidelity \cite{MU49}, especially when the number of physical qubits involved is large.
The steps of the simulations are as follows:
\begin{enumerate}
    \item Fix a depolarizing rate $p$.
    \item Randomly generate an error operator $E$ according to the probability distribution of a depolarizing channel.
    \item Compute the error syndrome and apply the correction operator and decoding operator.
    \item If there is no logical error after decoding, $E$ is correctable.
    \item Repeat steps 2 to 4 $N$ times.
    \item Output the channel fidelity as the number of correctable errors in the experiment divided by $N$.
\end{enumerate}
Two applications of the Monte Carlo method to the $[[11,1,5]]$ and $[[24,1,8]]$ codes obtained by MAGMA \cite{MAGMA}
are shown in Fig. \ref{fig:mc_11_1_5} and Fig. \ref{fig:mc_13_1_8_11}.
In Fig. \ref{fig:mc_11_1_5}, the exact channel fidelity of  the $[[11,1,5]]$ code is also plotted, and it can be observed that the simulations quickly converge to the exact channel fidelity.
On the other hand, in Fig. \ref{fig:mc_13_1_8_11}, only a lower bound of the channel fidelity  of the $[[24,1,8]]$ code is given, since computation of the channel fidelity is difficult ($4^{23}$ terms in the formula).
However, Monte Carlo simulations also converge quickly from $N=10^4$ to $10^6$ points, which are much less than $4^{23}\backsimeq 7\times 10^{13}$.

\section{Performance analysis} \label{sec:comparison}

\begin{figure*}
\begin{center}
\includegraphics[height=5.5cm,width=8.5cm]{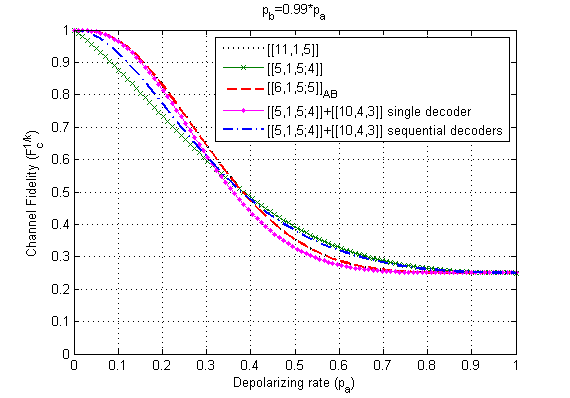}
\includegraphics[height=5.5cm,width=8.5cm]{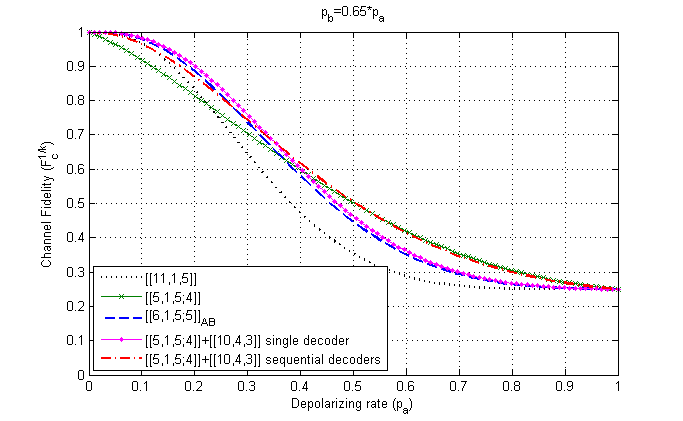}
\includegraphics[height=5.5cm, width=8.5cm]{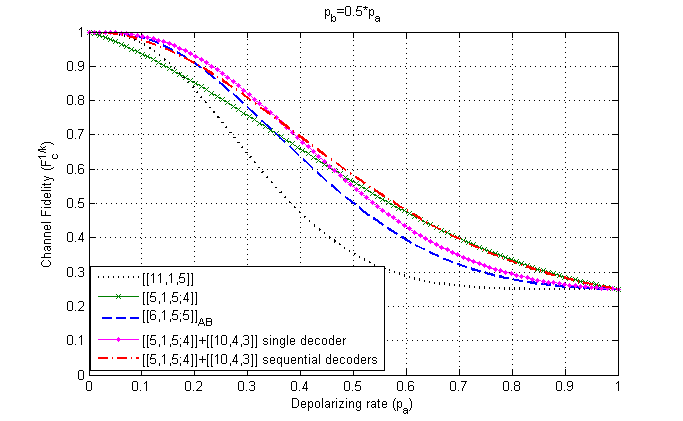}
\includegraphics[height=5.5cm, width=8.5cm]{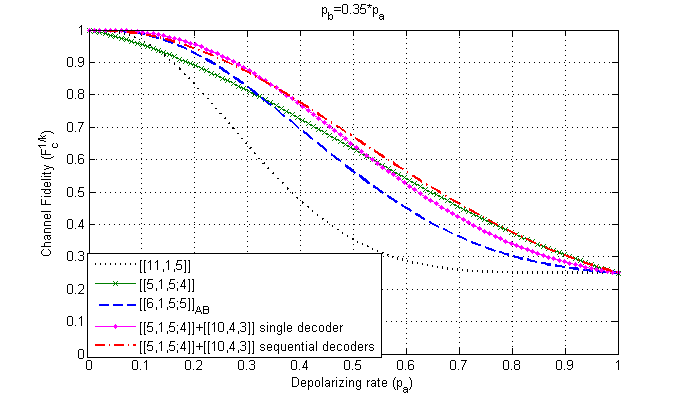}
\includegraphics[height=5.5cm, width=8.5cm]{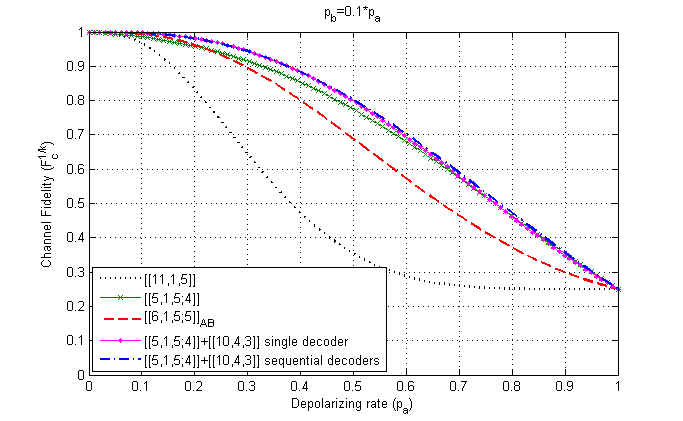}
\includegraphics[height=5.5cm, width=8.5cm]{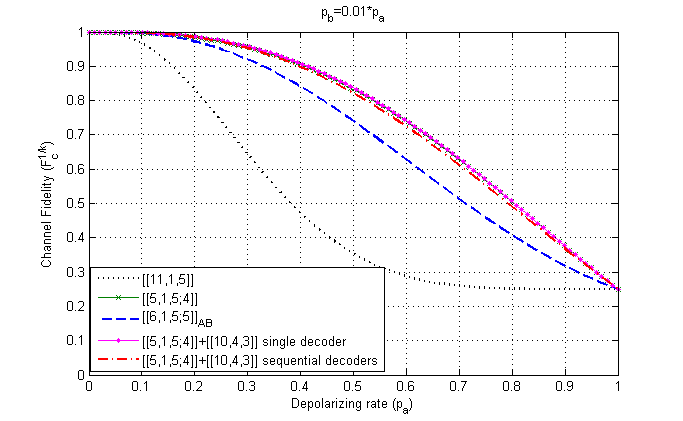}
\caption{(Color online) Channel fidelity of different quantum codes for $p_b=0.99,\ 0.65,\ 0.5,\ 0.35,\ 0.1,\ 0.01\ p_a.$
} \label{fig:channel fidelity}

\end{center}
\end{figure*}

In this section, we compare the channel fidelity of  the $[[5,1,5;4]]+[[10,4,3]]$ quantum code with other quantum codes.
A comparable standard stabilizer code with $n=15$ and $k=1$ is a $[[15,1,5]]$ stabilizer code,
while the smallest quantum code with $d=5$ is a $[[11,1,5]]$ stabilizer code \cite{Grassl}.
By Theorem \ref{thm:AB_code}, we can obtain a $[[6,1,5;5]]_{AB}$ EAQEC code from the $[[11,1,5]]$.
The $[[5,1,5;4]]$ EAQEC code is also shown as a reference.
All the quantum codes above encode $k=1$ information qubit and hence can be compared with no ambiguity.
The channel fidelity of these quantum codes are plotted in Fig. \ref{fig:channel fidelity}.
For simplicity, the $[[5,1,5;4]]+[[10,4,3]]$ quantum code using a single decoder is treated as a standard stabilizer code,
which means that the weight distribution of the syndrome representatives is like a triangle as in Fig. \ref{fig:weight_distribution_syndrome_representatives}.


\begin{figure}
\begin{center}
\includegraphics[height=6.5cm]
{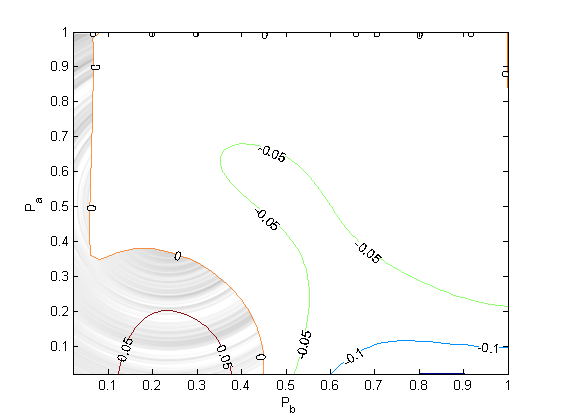}
{\caption{(Color online) Contour plot of the difference ($F_{single}-F_{seq}$) between  decoding methods for the $[[5,1,5;4]]+[[10,4,3]]$ quantum code.
The shaded region is where $F_{single}-F_{seq}>0$.
} \label{fig:two_enc vs one}}
\end{center}
\end{figure}

When $p_b=0.99 p_a$, the $[[6,1,5;5]]_{AB}$ EAQEC code is the best code for $p_a<~0.38$,
 and  $[[5,1,5;4]]+[[10,4,3]]$ quantum code with sequential decoders and the $[[5,1,5;4]]$ EAQEC code have better performance for  $p_a>~0.38$.
 The $[[5,1,5;4]]+[[10,4,3]]$ quantum code with a single decoder has a comparable fidelity to the $[[6,1,5;5]]_{AB}$ EAQEC code.

 As the ratio of $p_b$ to $p_a$ decreases, the performance of the $[[5,1,5;4]]+[[10,4,3]]$ quantum code and  the $[[6,1,5;5]]_{AB}$, $[[5,1,5;4]]$ EAQEC codes get better.
 When $p_b=0.5 p_a$, the $[[5,1,5;4]]+[[10,4,3]]$ quantum code, using either decoding method, performs better than the $[[6,1,5;5]]_{AB}$ EAQEC code and the $[[11,1,5]]$ stabilizer code.
When the ratio of $p_b$ to $p_a$ is below $0.35$ in the last two cases, the $[[5,1,5;4]]+[[10,4,3]]$ quantum code is the best choice.
The $[[5,1,5;4]]$ EAQEC code beats the $[[11,1,5]]$ stabilizer code when $p_b=0.01p_a$ as expected.
From these plots, the quantum codes that have the best performance are the  $[[6,1,5;5]]_{AB}$ EAQEC code or the $[[5,1,5;4]]+[[10,4,3]]$ quantum code.

Fig. \ref{fig:two_enc vs one} plots the
difference ($F_{single}-F_{seq}$) between the fidelity of a single decoder ($F_{single}$) and the fidelity of two sequential decoders  ($F_{seq}$) of the $[[5,1,5;4]]+[[10,4,3]]$ quantum code.
For low noise rates sequential decoding is not as good; however, it is easier to be implemented.
Note that, in the simulations, the syndrome representatives using a single decoder are chosen for the same error probabilities on Alice's and Bob's qubits.
The channel fidelity of the combination using a single decoder will be better if we optimize it over the choices of the syndrome representatives according to the different error probabilities.

\section{EAQEC and Entanglement Distillation}
\label{sec:Entanglement Distillation}

Up to now, we have assumed that Alice and Bob have perfect ebits before the communication, and any noise results from storage errors.
In a more general situation, the exchange of perfect maximally-entangled states might not be possible,
and entanglement distillation is needed.
Our next  direction is to find good strategies against this problem.

We first introduce the entanglement distillation protocol.
Suppose Alice has the ability to prepare $n$ pairs of maximally-entangled states (Bell states)
 \[ \ket{\Phi_+}^{\otimes m}=\left(\frac{1}{\sqrt{2}}(\ket{00}+\ket{11})\right)^{\otimes m} = \frac{1}{\sqrt{2^m}}\sum_{i=0}^{2^m-1} \ket{i}^{A}\ket{i}^{B},\]
where $i$ is the binary representation of the numbers between $0$ and $2^m-1$.
The state $\ket{\Phi_+}^{\otimes m}$ has the following property.
\bl
For any operator $M$ on an $m$-qubit maximally entangled state,
\[M^A \otimes I^B \ket{\Phi_+}^{\otimes m} = I^A \otimes (M^T)^B \ket{\Phi_+}^{\otimes m}, \]
where $M^T$ is the transpose of $M$.
\el
Suppose Alice uses an $[[m,c]]$ stabilizer code defined by a stabilizer group  $\mathcal{S}$ with generators $f_1, \cdots, f_{m-c}$, and $T$ is a set of syndrome representatives corresponding to $\mathcal{S}$.
Let $U$ be  a Clifford encoder of the stabilizer code, and  $\ket{j}_L$ for $j=0, \cdots, 2^c-1$ be the logical states.
The encoded $\ket{\Phi_+}^{\otimes c}$ is
\[
    \ket{\Phi_+}_L^{\otimes c}= U^A\otimes U^B\ket{\Phi_+}^{\otimes c}=\frac{1}{\sqrt{2^c}}\sum_{j=0}^{2^c-1}\ket{j}_L^A  \ket{j}_L^B.
\]
We know that $\bra{i}_L E_{s_1}E_{s_2}\ket{j}_L= \delta_{s_1,s_2} \delta_{i,j}$ and $\{E_s\ket{i}_L \}$ is a set of orthonormal basis vectors of $\mathcal{H}^{\otimes n}$.
In the case of $UU^T=I$, we have
\begin{align*}
\ket{\Phi_+}^{\otimes m}&= (UU^T)^A\otimes I^B \ket{\Phi_+}^{\otimes m}=  U^A \otimes U^B \ket{\Phi_+}^{\otimes m}\\
&=  \frac{1}{\sqrt{2^m}}\sum_{i=0}^{2^m-1}U^A \otimes U^B \ket{i}^{A}\ket{i}^{B}\\
&= \frac{1}{\sqrt{2^m}} \sum_{E_s\in T}  E_s^AE_s^B  \left(\sum_{j=0}^{2^c-1} \ket{j}_L^A  \ket{j}_L^B \right).
\end{align*}
If $UU^T\neq I$, Alice applies the operator $UU^T$ on half of the ebits.
From Wilde's encoding circuit algorithm \cite{wilde-2008}, an encoding operator can be implemented by a series of Hadamard gates, CNOT gates, SWAP gates, and phase gates.
If phase gates are not used in the circuit, the circuit will satisfy  $UU^T=I$.

Alice sends half of the ebits to Bob through a noisy channel  $\mathcal{N}_C$ with depolarizing rate equal to $p_c$.
The corrupted state is
\begin{align*}
E^B_i\ket{\Phi_+}^{\otimes m}= \frac{1}{\sqrt{2^m}}\sum_{j=0}^{2^c-1}\sum_{E_s\in T} E_s^A E^B_i E_s^B\ket{j}_L^A  \ket{j}_L^B.
\end{align*}
After performing a syndrome measurement, Alice obtains a syndrome $a$, which is a binary $(m-c)$-tuple,
and she sends $a$ to Bob through a noiseless classical channel.
Now the state is
\begin{align*}
 E_a^A E^B_i E_a^B \ket{\Phi_+}_L^{\otimes c}=  \frac{1}{\sqrt{2^c}}\sum_{j=0}^{2^c-1} E_a^A E^B_i E_a^B\ket{j}_L^A  \ket{j}_L^B,
\end{align*}
where $E_a$ is the syndrome representative of $a$.
Let $s(i)$ be the error syndrome of $E_i$.
Bob measures the stabilizer generators  $f_1, \cdots, f_{m-c}$ and obtains the syndrome \[b= a + s(i).\]
The error syndrome $s(i)$ can be retrieved by $s(i)=a+b$.
He applies the correction operator $E_{s(i)}$ and obtains the state
\begin{align*}
 E_a^A E_{s(i)}^B E^B_i E_a^B \ket{\Phi_+}_L^{\otimes c}.
\end{align*}
Finally, they restore the state to the standard encoded state  $\ket{\Phi_+}_L^{\otimes c}$
by applying the operator $E_a \otimes E_a$, followed
by the decoding circuit $U^{\dag}\otimes U^{\dag}$ and obtain the state
\[
 \ket{\Phi_+}^{\otimes c} \otimes \ket{0}^{m-c}\otimes \ket{0}^{m-c},
\]
if $E_i$ is correctable.
This is called an $[[m,c]]$ entanglement distillation protocol.
Let $\mathcal{E}$ denote the combination of the above processes.
The entanglement fidelity \cite{Schumacher96} of this protocol is
\[
F(\ket{\Phi_+}^{\otimes c},\mathcal{E} )= \bra{\Phi_+}^{\otimes c}\mathcal{E}\left(  \ket{\Phi_+}^{\otimes c}\bra{\Phi_+}^{\otimes c} \right) \ket{\Phi_+}^{\otimes c}.
\]
The entanglement fidelity of the $[[m,c]]$ entanglement distillation protocol  when the channel $\mathcal{N}_C$ is the depolarizing channel
is just the channel fidelity of the corresponding $[[m,c]]$ stabilizer code  over the depolarizing channel $\mathcal{N}_C$.

\begin{figure}
\begin{center}
\includegraphics[height=6.5cm,width=9cm]{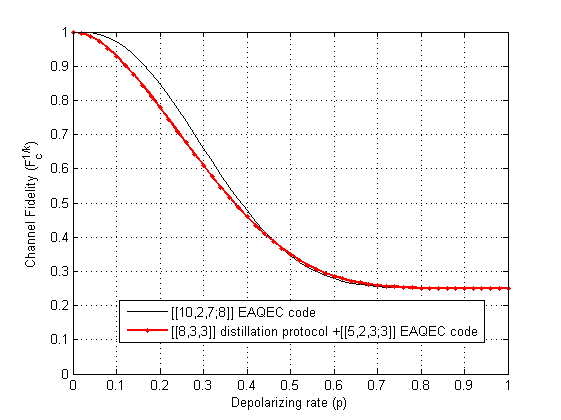}
{\caption{(Color online) Comparison of the $[[8,3,3]]$ distillation protocol plus $[[5,2,3;3]]$ EAQEC code, and the $[[10,2,7;8]]_{EA}$ EAQEC code without distillation.
The performance of the $[[10,2,7;8]]_{AB}$ EAQEC code is better for $p_a<0.45$.} \label{fig:distillation10278}}
\end{center}
\end{figure}

After performing an $[[m,c]]$ entanglement distillation protocol that produces $c$  ebits with some entanglement fidelity, Alice can use an $[[n,k,d;c]]$ EAQEC code to send $k$ information qubits to Bob.  Assume $p_b=0$ and  $p_c=p_a$ for simplicity.
The channel fidelity from the combination of an entanglement distillation protocol and an EAQEC code is similar to that of a combination code using two sequential decoders.  On the other hand, Alice could directly use an $[[n',k,d';e]]$ EAQEC code with the imperfect ebits to send $k$ information qubits to Bob without entanglement distillation.  The channel fidelity of this process is just the channel fidelity of the EAQEC code.
\be
Alice and Bob use an $[[8,3,3]]$ entanglement distillation protocol to produce 3 perfect ebits from 8 noisy ebits.  Then Alice can use an $[[5,2,3;3]]$ EAQEC code to send quantum information to Bob.  Or she can instead use a $[[10,2,7;8]]$ EAQEC code with the 8 corrupted ebits.  Comparison of the channel fidelity of these two schemes is shown in Fig. \ref{fig:distillation10278}.
\ee

\begin{figure}
\begin{center}
\includegraphics[height=6cm,width=9cm]{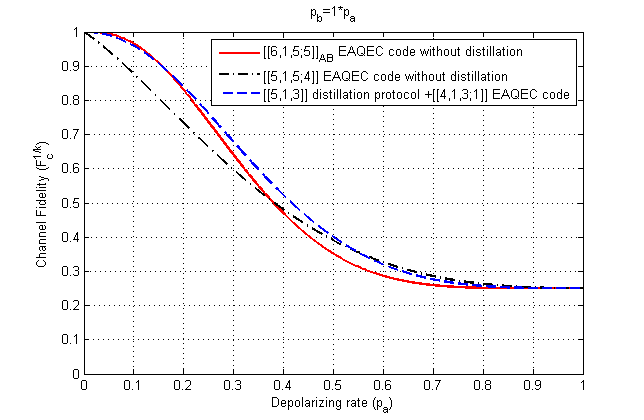}  {\caption{(Color online) Comparison of the $[[5,1,3]]$ distillation protocol plus $[[4,1,3;1]]$ EAQEC code, and the $[[6,1,5;5]]_{EA}$ EAQEC code without distillation.
The performance of the $[[6,1,5;5]]_{AB}$ EAQEC code is slightly better for $p_a<0.11$.}
\label{fig:distillation6155}}
\end{center}

\end{figure}

\be
Alice and Bob use a $[[5,1,3]]$ entanglement distillation protocol to produce 1 perfect ebit from 5 noisy ebits.
Then Alice can use a $[[4,1,3;1]]$ EAQEC code to send quantum information to Bob.
Or she can instead use a $[[6,1,5;5]]_{AB}$ EAQEC code with the 5 noisy ebits.
Comparison of the channel fidelity of these two schemes is shown in Fig. \ref{fig:distillation6155}.

\ee

%

\begin{figure}
\begin{center}
\includegraphics[height=6cm,width=8.5cm]{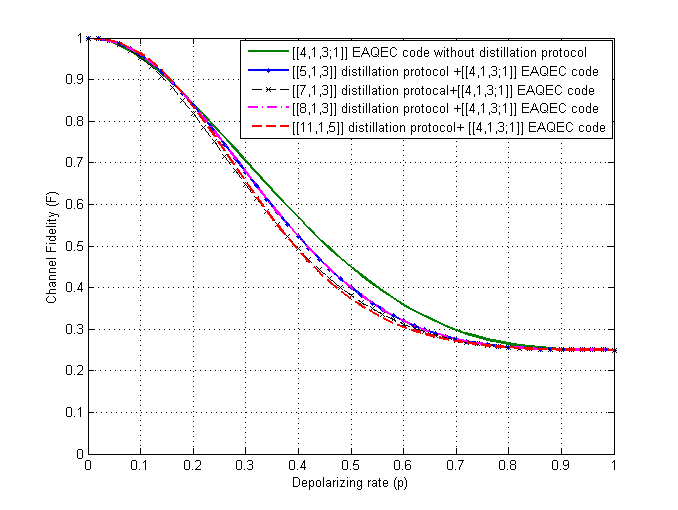}  \label{fig:diff_distillation_same_eaqec}
{\caption{(Color online) Comparison of different distillation protocols  with the same $[[4,1,3;1]]$ EAQEC code.
 }\label{fig:distillation_4131}}
\end{center}
\end{figure}

In Fig. \ref{fig:distillation_4131}, we plot the channel fidelity of four distillation protocols with the same $[[4,1,3;1]]$ EAQEC code.
The channel fidelities of the 4 schemes with distillation protocols are better than the one without distillation protocol for $p<0.1$, but the difference is small. The $[[4,1,3;1]]$ EAQEC code without distillation dominates the performance for higher $p$.

In Fig. \ref{fig:distillation_k1}, we plot the channel fidelity of several coding schemes that encode $k=1$ information qubit.
It can be observed that the EAQEC codes without distillation protocols have better performance for $p<0.2$.

\begin{figure}
\begin{center}
\includegraphics[height=6.5cm,width=8.5cm]{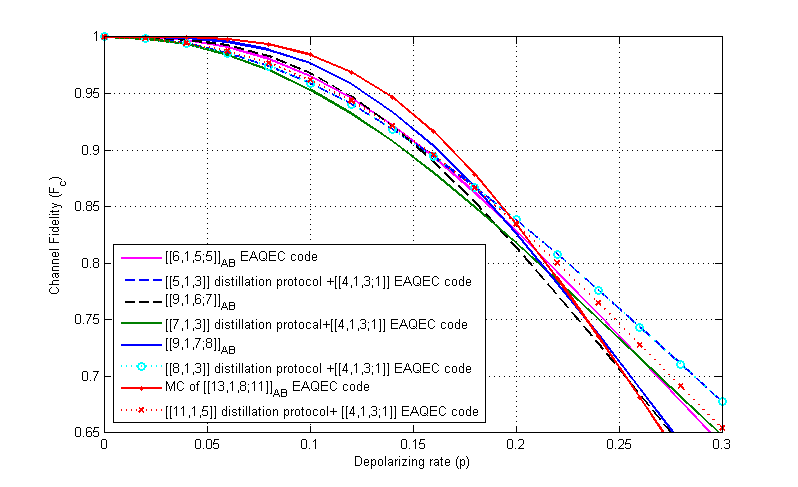}  \label{fig:distillation_k1}
{\caption{(Color online) Comparison of  different combinations of distillation protocols and EAQEC codes that encode one information qubit.
 }\label{fig:distillation_k1}}
\end{center}
\end{figure}

\section{Discussion} \label{sec:discussion}
In this paper, we proposed two coding schemes for EAQEC codes when the ebits of the receiver suffer errors.  In the first case we assume the ebits suffer storage errors. EAQEC codes that are equivalent to standard stabilizer codes have better performance than their corresponding stabilizer codes.  Several such EAQEC codes are found to achieve the linear programming bound, and hence are optimal.  However, as long as the storage error rate is small enough, a quantum code with two encoders performs well if we start with an EAQEC code that is not equivalent to any standard stabilizer codes.  We may choose the best quantum code according to the noise channel rate and the storage error rate in real situations.


Any (nondegenerate) standard stabilizer code can be transformed into an EAQEC code by  Theorem \ref{thm:AB_code}.  Families of quantum codes, such as quantum Reed-Muller codes \cite{Ste98}, quantum BCH codes \cite{GB99,AKS06}, quantum cyclic codes \cite{CohenLitsyn99,Lin04}, can be transformed into families of EAQEC codes.  It is possible to construct EAQEC codes with a large number of information qubits but a small number of ebits that outperform standard codes.

We developed a formula for the channel fidelity over the depolarizing channel,
 and used it to evaluate the performance of a variety of quantum codes.  For large codes, the channel fidelity cannot be calculated exactly, but can be lower-bounded or approximated by Monte Carlo simulations.  A similar formula for the channel fidelity  can be developed for other channels with only Pauli errors.

We also compared EAQEC codes combined with entanglement distillation protocols to EAQEC codes designed to tolerate noisy ebits.  It seems that EAQEC codes that can correct errors on both the qubits of sender and receiver can have better performance than the codes combined with an entanglement distillation protocol, at least for modest noise rates.  For particular combinations of error rates and applications, it should be possible to optimize the choice of code to maximize the fidelity.  This optimization is the subject of ongoing research.

\begin{acknowledgments}
TAB and CYL acknowledge useful conversations with Igor Devetak and Mark Wilde.  This work was supported in part by NSF Grants CCF-0448658 and CCF-0830801.
\end{acknowledgments}

\bibliography{qecc_APS}

\end{document}